\documentclass[12pt,a4paper]{article}
\usepackage{axodraw}

\makeatletter

\def\paragraph{\@startsection{paragraph}{4}{\z@}{+2.00ex plus
 +1ex minus +.2ex}{1.5ex plus .2ex}{\it\normalsize}}

\def\section{\@startsection {section}{1}{\z@}{+3.0ex plus +1ex minus
  +.2ex}{2.3ex plus .2ex}{\normalsize\bf\boldmath}}
\def\subsection{\@startsection{subsection}{2}{\z@}{+2.5ex plus +1ex
minus +.2ex}{1.5ex plus .2ex}{\normalsize\bf\boldmath}}
\def\subsubsection{\@startsection{subsubsection}{3}{\z@}{+3.25ex plus
 +1ex minus +.2ex}{1.5ex plus .2ex}{\normalsize\it}}

\expandafter\ifx\csname mathrm\endcsname\relax\def\mathrm#1{{\rm #1}}\fi

\newcounter{saveeqn}

\@addtoreset{equation}{section}

\newcount\@tempcntc
\def\@citex[#1]#2{\if@filesw\immediate\write\@auxout{\string\citation{#2}}\fi
  \@tempcnta\z@\@tempcntb\m@ne\def\@citea{}\@cite{\@for\@citeb:=#2\do
    {\@ifundefined
       {b@\@citeb}{\@citeo\@tempcntb\m@ne\@citea
        \def\@citea{,\penalty\@m\ }{\bf ?}\@warning
       {Citation `\@citeb' on page \thepage \space undefined}}%
    {\setbox\z@\hbox{\global\@tempcntc0\csname
b@\@citeb\endcsname\relax}%
     \ifnum\@tempcntc=\z@ \@citeo\@tempcntb\m@ne
       \@citea\def\@citea{,\penalty\@m}
       \hbox{\csname b@\@citeb\endcsname}%
     \else
      \advance\@tempcntb\@ne
      \ifnum\@tempcntb=\@tempcntc
      \else\advance\@tempcntb\m@ne\@citeo
      \@tempcnta\@tempcntc\@tempcntb\@tempcntc\fi\fi}}\@citeo}{#1}}

\def\@citeo{\ifnum\@tempcnta>\@tempcntb\else\@citea
  \def\@citea{,\penalty\@m}%
  \ifnum\@tempcnta=\@tempcntb\the\@tempcnta\else
   {\advance\@tempcnta\@ne\ifnum\@tempcnta=\@tempcntb \else
\def\@citea{--}\fi
    \advance\@tempcnta\m@ne\the\@tempcnta\@citea\the\@tempcntb}\fi\fi}

\def\beq{\begin{equation}}
\def\eeq{\end{equation}}
\def\beqar{\begin{eqnarray}}
\def\eeqar{\end{eqnarray}}
\def\barr#1{\begin{array}{#1}}
\def\earr{\end{array}}
\def\bfi{\begin{figure}}
\def\efi{\end{figure}}
\def\btab{\begin{table}}
\def\etab{\end{table}}
\def\bce{\begin{center}}
\def\ece{\end{center}}
\def\nn{\nonumber}
\def\nl{\nonumber\\}

\def\al{\alpha}

\def\de{\delta}
\def\la{\lambda}
\def\si{\sigma}
\def\Si{\Sigma}

\def\refeq#1{\mbox{(\ref{#1})}}
\def\reffi#1{\mbox{Figure~\ref{#1}}}
\def\reffis#1{\mbox{Figures~\ref{#1}}}

\def\refse#1{\mbox{Section~\ref{#1}}}
\def\refses#1{\mbox{Sections~\ref{#1}}}
\def\refapp#1{\mbox{App.~\ref{#1}}}
\def\citere#1{\mbox{Ref.~\cite{#1}}}
\def\citeres#1{\mbox{Refs.~\cite{#1}}}

\newcommand{\TeV}{\unskip\,\mathrm{TeV}}
\newcommand{\GeV}{\unskip\,\mathrm{GeV}}

\newcommand{\ri}{{\mathrm{i}}}
\newcommand{\rd}{{\mathrm{d}}}
\newcommand{\rS}{{\mathrm{S}}}
\newcommand{\rR}{{\mathrm{R}}}
\newcommand{\rT}{{\mathrm{T}}}
\newcommand{\rL}{{\mathrm{L}}}

\renewcommand{\O}{{\cal O}}
\newcommand{\M}{{\cal{M}}}

\def\mathswitchr#1{\relax\ifmmode{\mathrm{#1}}\else$\mathrm{#1}$\fi}
\newcommand{\PW}{\mathswitchr W}
\newcommand{\PB}{\mathswitchr B}
\newcommand{\PZ}{\mathswitchr Z}
\newcommand{\PA}{\mathswitchr A}
\newcommand{\PH}{\mathswitchr H}
\newcommand{\Pf}{\mathswitch f}
\newcommand{\Pfbar}{\mathswitch \bar f}
\newcommand{\Pe}{\mathswitchr e}
\newcommand{\Pd}{\mathswitchr d}
\newcommand{\Pu}{\mathswitchr u}
\newcommand{\Ps}{\mathswitchr s}
\newcommand{\Pc}{\mathswitchr c}
\newcommand{\Pb}{\mathswitchr b}
\newcommand{\Pt}{\mathswitchr t}
\newcommand{\Pep}{\mathswitchr {e^+}}
\newcommand{\Pem}{\mathswitchr {e^-}}
\newcommand{\PWp}{\mathswitchr {W^+}}
\newcommand{\PWm}{\mathswitchr {W^-}}
\newcommand{\PWpm}{\mathswitchr {W^\pm}}

\def\mathswitch#1{\relax\ifmmode#1\else$#1$\fi}
\newcommand{\MW}{\mathswitch {M_\PW}}
\newcommand{\MZ}{\mathswitch {M_\PZ}}
\newcommand{\MH}{\mathswitch {M_\PH}}

\newcommand{\Mt}{\mathswitch {m_\Pt}}

\newcommand{\thw}{\mathswitch {\theta_\mathrm{w}}}
\newcommand{\cw}{\mathswitch {c_\mathrm{w}}}
\newcommand{\sw}{\mathswitch {s_\mathrm{w}}}

\newcommand{\NCf}{\mathswitch {N_{\mathrm{C}}^f}}
\newcommand{\NCt}{\mathswitch {N_{\mathrm{C}}^{\Pt}}}

\newcommand{\Tr}{\mathop{\mathrm{Tr}}\nolimits}
\newcommand{\SU}{\mathrm{SU}}

\newcommand{\SUtwo}{\mathrm{SU(2)}}
\newcommand{\Uone}{\mathrm{U}(1)}

\def\ie{i.e.\ }

\def\cf{cf.\ }

\newcommand{\elm}{{\mathrm{em}}}
\newcommand{\ew}{{\mathrm{ew}}}
\newcommand{\sew}{{\mathrm{ew}}}

\newcommand{\coll}{{\mathrm{coll}}}

\newcommand{\symm}{{\mathrm{symm}}}
\newcommand{\asymm}{{\mathrm{asymm}}}

\newcommand{\SC}{{\mathrm{LSC}}}
\renewcommand{\SS}{{\mathrm{SSC}}}
\newcommand{\cc}{{\mathrm{C}}}

\newcommand{\pre}{{\mathrm{PR}}}
\newcommand{\Yuk}{{\mathrm{Yuk}}}

\newcommand{\bew}{b^{\ew}}
\newcommand{\besw}{\tilde{b}^{\ew}}

\newcommand{\cew}{C^{\ew}}
\newcommand{\csew}{\tilde{C}^{\ew}}
\newcommand{\dew}{D^{\ew}}
\newcommand{\dsew}{\tilde{D}^{\ew}}
\newcommand{\sNB}{\tilde{\NB}}
\newcommand{\NB}{N}
\newcommand{\GB}{V}

\newcommand{\ls}{l(s)}
\newcommand{\lu}{l(\mu^2)}
\newcommand{\lrM}{l(r_{kl},M^2)}

\newcommand{\lsM}{l(s,M^2)}
\newcommand{\lsW}{l(s,\MW^2)}

\newcommand{\lsl}{l_{\cc}}
\newcommand{\lpr}{l_{\pre}}
\newcommand{\lYuk}{l_{\Yuk}}
\newcommand{\lZ}{l_{\PZ}}
\newcommand{\lWf}{l(\MW^2,m_f^2)}
\newcommand{\lWfsi}{l(\MW^2,m_{f_\si}^2)}

\newcommand{\lWla}{l(\MW^2,\la^2)}
\newcommand{\lWfsii}{l(\MW^2,m_{f_{\si,i}}^2)}

\newcommand{\lWa}{l(\MW^2,M_{\GB_a}^2)}
\newcommand{\lWZ}{l(\MW^2,\MZ^2)}
\newcommand{\ltW}{l(\Mt^2,\MW^2)}
\newcommand{\lHW}{l(\MH^2,\MW^2)}
\newcommand{\lemf}{l^\elm(m_f^2)}
\newcommand{\lemftau}{l^\elm(m_{f_\tau}^2)}
\newcommand{\lemfsi}{l^\elm(m_{f_\si}^2)}

\newcommand{\lemphi}{l^\elm(m_{\varphi}^2)}
\newcommand{\leme}{l^\elm(m_\Pe^2)}
\newcommand{\lemW}{l^\elm(\MW^2)}
\newcommand{\Ls}{L(s)}
\newcommand{\LrM}{L(|r_{kl}|,M^2)}
\newcommand{\Lrs}{L(|r_{kl}|,s)}
\newcommand{\LrMa}{L(|r_{kl}|,M_{\GB_a}^2)}
\newcommand{\LsM}{L(s,M^2)}
\newcommand{\LsW}{L(s,\MW^2)}

\newcommand{\Lkla}{L(m_k^2,\la^2)}

\newcommand{\LWla}{L(\MW^2,\lambda^2)}
\newcommand{\Lemk}{L^\elm(s,\lambda^2,m_k^2)}
\newcommand{\Lemphi}{L^\elm(s,\lambda^2,m_\varphi^2)}

\newcommand{\Lemftau}{L^\elm(s,\lambda^2,m_{f_\tau}^2)}
\newcommand{\Leme}{L^\elm(s,\lambda^2,m_e^2)}

\newcommand{\lrs}{\log{\frac{|r_{kl}|}{s}}}

\newcommand{\ltu}{\log{\frac{t}{u}}}
\newcommand{\lts}{\log{\frac{|t|}{s}}}
\newcommand{\lus}{\log{\frac{|u|}{s}}}

\def\draftdate{\relax}
\def\mda{\relax}
\def\mua{\relax}
\def\mla{\relax}
\def\draft{
\def\thtystars{******************************}
\def\sixtystars{\thtystars\thtystars}
\typeout{}
\typeout{\sixtystars**}
\typeout{* Draft mode!
         For final version remove \protect\draft\space in source file *}
\typeout{\sixtystars**}
\typeout{}
\def\draftdate{\today}
\def\mua{\marginpar[\boldmath\hfil$\uparrow$]%
                   {\boldmath$\uparrow$\hfil}%
                    \typeout{marginpar: $\uparrow$}\ignorespaces}
\def\mda{\marginpar[\boldmath\hfil$\downarrow$]%
                   {\boldmath$\downarrow$\hfil}%
                    \typeout{marginpar: $\downarrow$}\ignorespaces}
\def\mla{\marginpar[\boldmath\hfil$\rightarrow$]%
                   {\boldmath$\leftarrow $\hfil}%
                    \typeout{marginpar: $\leftrightarrow$}\ignorespaces}
\def\Mua{\marginpar[\boldmath\hfil$\Uparrow$]%
                   {\boldmath$\Uparrow$\hfil}%
                    \typeout{marginpar: $\uparrow$}\ignorespaces}
\def\Mda{\marginpar[\boldmath\hfil$\Downarrow$]%
                   {\boldmath$\Downarrow$\hfil}%
                    \typeout{marginpar: $\downarrow$}\ignorespaces}
\def\Mla{\marginpar[\boldmath\hfil$\Rightarrow$]%
                   {\boldmath$\Leftarrow $\hfil}%
                    \typeout{marginpar: $\leftrightarrow$}\ignorespaces}
\def\sua{\Mua}
\def\sda{\Mda}
\def\sla{\Mla}
\overfullrule 5pt
\oddsidemargin -15mm
\marginparwidth 29mm
}

\begin{document}
\thispagestyle{empty}
\def\thefootnote{\fnsymbol{footnote}}
\setcounter{footnote}{1}
\null
\draftdate\hfill  PSI-PR-00-15\\
\strut\hfill ZU-TH-17-00\\
\strut\hfill hep-ph/0010201
\vskip 0cm
\vfill
\begin{center}
{\Large \bf
One-loop leading logarithms\\ in electroweak radiative corrections\\ I. Results
\par} \vskip 2.5em
{\large
{\sc A.~Denner%
}\\[1ex]
{\normalsize \it Paul Scherrer Institut\\
CH-5232 Villigen PSI, Switzerland}\\[2ex]
{\sc S.~Pozzorini%
}\\[1ex]
{\normalsize \it
Institute of Theoretical Physics\\ University of Z\"urich, Switzerland \\[2ex]and \\[2ex]
{\normalsize \it Paul Scherrer Institut\\
CH-5232 Villigen PSI, Switzerland}\\[2ex]
}}

\par \vskip 1em
\end{center}
\par
\vskip .0cm \vfill {\bf Abstract:} \par 
We present results for the complete one-loop electroweak logarithmic
corrections for general processes at high energies and fixed angles.
Our results are applicable to arbitrary matrix elements that are not
mass-suppressed.  We give explicit results for 4-fermion processes and
gauge-boson-pair production in $\Pep\Pem$ annihilation.
\par
\vskip 1cm
\noindent
October 2000 
\par
\null
\setcounter{page}{0}
\clearpage
\def\thefootnote{\arabic{footnote}}
\setcounter{footnote}{0}

\section{Introduction}

In the LEP regime, at energies $\sqrt{s}\sim \MZ$, electroweak
radiative corrections 
are dominated by large electromagnetic effects from initial-state
radiation, by the contributions of the running electromagentic
coupling, and by the corrections associated with the $\rho$ parameter,
typically of the order $10\%$.  Future colliders, such as the LHC 
\cite{cernreport} or an $\Pep\Pem$ linear collider (LC) \cite{LC1},
will explore a new energy range, $\sqrt{s}\gg \MZ$.  It is known since
many years (see, for instance, \citeres{Kuroda,eeWWhe}) that above the
electroweak scale the structure of the leading electroweak corrections
changes and double logarithms of Sudakov type \cite{SUD}  as well
as single logarithms involving the ratio of the energy to the
electroweak scale become dominating.
These logarithms arise from virtual (or real) gauge bosons emitted by
the initial and final-state particles. They correspond to the
well-known soft and collinear singularities observed in theories with
massless gauge bosons.

In massless gauge theories such as QED and QCD, the soft and collinear logarithms in the virtual corrections  are singular and 
have to be cancelled by adding the
contribution of real gauge-boson radiation.  In the electroweak
theory, the masses of the weak gauge bosons, $\PZ$ and $\PW$, provide
a physical cutoff, and the massive gauge bosons can be detected
as distinguished particles.  Unlike for the photon, real $\PZ$ and
$\PW$ bremsstrahlung need not be included, and the large logarithms
originating from virtual corrections are of physical significance.

The typical size of double-logarithmic (DL) and single-logarithmic
(SL) corrections is given by
\beq
\frac{\alpha}{4\pi \sw^2}\log^2{\frac{s}{\MW^2}}= 6.6 \%,\qquad \frac{\alpha}{4\pi \sw^2}\log{\frac{s}{\MW^2}}=1.3\%
\eeq
at $\sqrt{s}=1\TeV$ and increases with the energy. If the experimental
precision is at the few-percent level like at the LHC, both DL and SL
contributions have to be included at the one-loop level. In view of
the precision objectives of a LC, between the percent and
the permil level, besides the complete one-loop
corrections also two-loop DL effects have to be taken into account.
The DL contributions represent a leading and negative correction,
whereas the SL ones often have opposite sign, and are
referred to as subleading. The compensation between DL and SL
corrections can be quite important \cite{be1,Ku2}, and depending on
the process and the energy, the SL contribution can be even larger
than the DL one.%
\footnote{For instance in $e^+e^-\rightarrow \mu^+
  \mu^-$ \cite{Ku2} at $\sqrt{s}=1 \TeV$ one has $+13.8\%$ for SL and
  $-9.6\%$ for DL corrections to the unpolarized cross
  section.}

Owing to this phenomenological relevance, the infrared (IR) 
structure of the
electroweak theory is receiving increasing interest recently.  The
one-loop structure and the origin of the DL corrections have been
discussed for $\Pep\Pem\to\Pf\Pfbar$ \cite{CC0,Ku1} and are by
now well established. Recipes for the resummation of the DL
corrections have been developed
\cite{CC1,Ku1,Ku2,Fa00} and explicit calculations of the leading DL
corrections for the processes $g\to\Pf\Pfbar$ and
$\Pep\Pem\to\Pf\Pfbar$ have been performed \cite{Me2,Be00,Ho00}.
On the other hand, for the SL corrections complete one-loop
calculations are only available for 4-fermion
neutral-current processes \cite{be1,Ku2} and \PW-pair production 
\cite{eeWWhe}. The subleading two-loop logarithmic corrections have
been evaluated for $\Pep\Pem\to\Pf\Pfbar$ in \citere{Ku2}. A general
recipe for a subclass of SL corrections to all orders has been proposed in
\citere{Me1}, based on the infrared-evolution-equation method.
 
In this paper, we present results for all DL and SL contributions to
the electroweak one-loop virtual corrections.  The results apply to
exclusive processes with arbitrary external states, including
transverse and longitudinal gauge bosons as well as Higgs fields.
Above the electroweak scale, the photon, $\PZ$- and $\PW$-boson loops
are most conveniently treated in a symmetric way, rather than split
into electromagnetic and weak parts \cite{CC1}; at the same time
special care has to be taken for the gap between the photon mass and
the weak scale $\MW$.  To this end we split the logarithms originating
from the electromagnetic and from the $\PZ$-boson loops into two
parts: the contributions of a fictitious heavy photon and a
$\PZ$-boson with mass $\MW$, which are added to the $\PW$-boson loops
resulting in the ``symmetric electroweak'' (sew) contribution, and the
remaining part originating from the difference between the photon or
$\PZ$-boson mass and the mass of the $\PW$-boson.
The large logarithms originating in the photon loops owing to the gap
between the electromagnetic and the weak scale are denoted as
``pure electromagnetic'' (em) contribution.

In contrast to predictions based on the unbroken phase, our results
are obtained from the high-energy limit of the broken phase, \ie with
calculations in the physical fields. In this way all features of the
electroweak theory are consistently implemented.%
\footnote{As observed in \citere{Ku1}, the Higgs mechanism is
  irrelevant for the IR structure at the DL level. This seems to be less
  clear at the SL level where, through self-energy contributions, mixing
  effects between gauge bosons and Goldstone bosons enter.}
Especially, the mixing and the mass gap between photons and $\PZ$ bosons
is well under control. Furthermore, the longitudinal components of
massive gauge bosons and the scalar fields are included 
as external states.

\subsection*{On the method}
We work within the 't~Hooft--Feynman gauge and use dimensional
regularization so that ultraviolet (UV) single logarithms depend on
the regularization scale $\mu$. Exploiting the $\mu$ independence of
the $S$ matrix, we choose $\mu^2=s$ so that the logarithms
$\log{(\mu^2/s)}$ 
related to the UV singularities are not enhanced, and only the
mass-singular logarithms $\log{(\mu^2/M^2)}$ or $\log{(s/M^2)}$ are
large. In order to be specific we fix the field renormalization
constants (FRCs) such that no extra wave-function renormalization
constants are required \cite{FortPhys}. For parameter renormalization
we adopt the on-shell scheme for
definiteness. This can be easily changed.  In this setup large
logarithms appear in the mass-singular loop diagrams as well as in the
coupling and field renormalization constants, and are distributed as
follows:
\begin{itemize}
\item The DL contributions originate from those one-loop diagrams
  where soft--collinear gauge bosons are exchanged between pairs of
  external legs. These double logarithms are obtained with the eikonal
  approximation.
\item The SL mass-singular contributions from loop diagrams originate
  from the emission of virtual collinear gauge bosons from external
  lines \cite{KLN}. These SL contributions are extracted from the loop
  diagrams in the collinear limit by means of Ward identities, and are
  found to factorize into the Born amplitude times ``collinear
  factors''. These are the main result of this article,
and a forthcoming publication \cite{DennPozz2} will be dedicated to a
detailed description of their calculation. 

\item The remaining SL contributions
originating from soft and collinear regions 
are contained in the FRCs.
\item The parameter renormalization constants, \ie the 
charge- and weak-mixing-angle renormalization constants as well as the
renormalization constants for Yukawa and scalar self couplings, involve the SL
contributions of UV origin. These are the leading logarithms that are 
controlled by the renormalization group.
\end{itemize}
The DL and SL mass-singular terms are extracted from loop diagrams by
setting all masses to zero in the numerators of the loop-integrals.
This approach is applicable only if
 no inverse powers of gauge-boson
masses are present in the Feynman rules. In the Feynman gauge this is
true except for the polarization vectors of longitudinal gauge bosons.
However, since we are only interested in the high-energy limit, we can
use the Goldstone-boson equivalence theorem \cite{et}
for processes involving longitudinal gauge-bosons taking into account
the correction factors from higher-order contributions \cite{etcorr}.

The paper is organized as follows: in \refse{se:defcon} we introduce
our basic definitions and conventions. The leading logarithms
originating from the soft--collinear region, from the soft or collinear
regions, and from parameter renormalization
are considered in \refses{se:soft-coll}, \ref{se:soft-or-coll}, and
\ref{se:ren}, respectively. In \refse{se:applicat} we discuss some
applications of our general results to simple specific processes.
Results for the electroweak logarithmic corrections to the production
of an arbitrary number of transverse gauge bosons in
fermion--antifermion annihilation are given in \refapp{app:transvRG}.
Finally, \refapp{app:representations} summarizes explicit results for
the various generic quantities appearing in our formulas.

\section{Definitions and conventions}
\label{se:defcon} 
We consider electroweak processes involving $n$ arbitrary external 
particles. As a convention, all these particles and their momenta
$p_k$ are assumed to be incoming, so that the process reads
\beq \label{process}
\varphi_{i_1}(p_1)\dots \varphi_{i_n}(p_n)\rightarrow 0.
\eeq
The particles (or antiparticles) $\varphi_{i_k}$ correspond to the
components of the various multiplets
$\varphi$ present in
the standard model. Chiral fermions and antifermions are represented
by $f^\kappa_\si$ and $\bar{f}^\kappa_\si$, respectively,  
with the chirality
$\kappa=\rR,\rL$ and the isospin indices $\si=\pm$. The gauge bosons
are denoted by $\GB_a=\PA,\PZ,\PWpm$, and can be transversely (T) or
longitudinally (L)
polarized. 
For neutral gauge bosons we use the symbol $\NB=\PA,\PZ$.
The components $\Phi_i$ of the scalar
doublet consist of the physical Higgs particle $\PH$ and the
unphysical Goldstone bosons $\chi,\phi^\pm$, which are used to
describe the longitudinally polarized massive gauge bosons
$\PZ_\rL$ and $\PW^\pm_\rL$ 
with help of the equivalence theorem.

The predictions for general  processes, 
\beq \label{22proc}
\varphi_{i_1}(p^{\mathrm{in}}_1)\ldots\varphi_{i_m}(p^{\mathrm{in}}_m)
\rightarrow 
\varphi_{j_1}(p^{\mathrm{out}}_1)\dots \varphi_{j_{n-m}}(p^{\mathrm{out}}_{n-m}),
\eeq
can be obtained  by crossing symmetry from our predictions for the
$n\rightarrow 0$ process 
\beq
\varphi_{i_1}(p^{\mathrm{in}}_1)\ldots\varphi_{i_m}(p^{\mathrm{in}}_m)
\bar{\varphi}_{j_1}(-p^{\mathrm{out}}_1)\dots\bar{\varphi}_{j_{n-m}}(-p^{\mathrm{out}}_{n-m})\rightarrow 0
\eeq 
where $\bar{\varphi}_i$ represents the charge conjugate of
$\varphi_{i}$. Thus, outgoing particles (antiparticles) are
substituted by incoming antiparticles (particles) and the
corresponding momenta are reversed.  These substitutions can be
directly applied to our results.

The couplings of the external fields  $\varphi_{i_k}$  to the gauge bosons
$\GB_a$ are denoted by $\ri eI^{\GB_a}(\varphi)$,
and correspond to the generators of infinitesimal global
$\SUtwo\times\Uone$ transformations of these fields,
\beq \label{generators}
\delta_{\GB_a} \varphi_{i}=
\ri e I^{\GB_a}_{\varphi_i\varphi_{i'}}(\varphi)\,\varphi_{i'}.
\eeq
To be precise, $\ri eI^{\GB_a}_{\varphi_i\varphi_{i'}}(\varphi)$ is the
coupling corresponding to the $\GB_a\bar{\varphi}_i\varphi_{i'}$ vertex,
where all fields are incoming.
The indices of the matrix
$I^{\GB_a}_{\varphi_i\varphi_{i'}}(\varphi)$ may be particles or
antiparticles, and charge conjugation of the identity
\refeq{generators} gives
\beq\label{eq:chargeconjcoup}
I^{\bar{\GB}_a}_{\bar{\varphi}_i\bar{\varphi}_{i'}}(\bar{\varphi})= -\left(I^{\GB_a}_{\varphi_i\varphi_{i'}}(\varphi)\right)^*.
\eeq
As a shorthand notation for those formulas where various fields
labelled by $k=1,\dots,n$ occur, the components $\varphi_{i_k}$ are
replaced by their indices $i_k$. For instance, the generators in
\refeq{generators} are denoted by $I^a_{i_ki'_k}(k)$.  A detailed
description of the generators and other group-theoretical operators is
given in \refapp{app:representations}, together with the explicit
values for various representations.

We consider the process \refeq{process} with all external momenta on
shell, $p_k^2=m_k^2$, and in the limit where all invariants are much
larger than the gauge-boson masses, in particular
\beq \label{Sudaklim} 
r_{kl}=(p_k+p_l)^2\sim 2p_kp_l \gg \MW^2.
\eeq
Note that this condition is not fulfilled if the 
cross section is dominated by resonances. 
We restrict
ourselves to Born matrix elements that are not mass-suppressed 
in this limit, and we calculate the virtual
one-loop corrections in
leading and subleading logarithmic approximation (LA), \ie we take
into account only enhanced DL and SL terms and omit non-enhanced
terms. The logarithmic  contributions are written in terms of  
\beq
\LrM:=\frac{\alpha}{4\pi}\log^2{\frac{r_{kl}}{M^2}},\qquad
\lrM:=\frac{\alpha}{4\pi}\log{\frac{r_{kl}}{M^2}},
\eeq
and depend on different invariants $r_{kl}$ and masses $M$, according
to the Feynman diagrams they originate from. In order to render the
results as symmetric as possible, we relate the energy-dependent part
of all large logarithms to the scales $\MW$ and $s$.  To this end, we
write all these logarithms in terms of
\beq \label{dslogs}
\Ls:=\LsW,\qquad
\ls:=\lsW,
\eeq
and logarithms of mass ratios and ratios of invariants.  The DL
contributions proportional to $\Ls$ and to $\ls\log(|r_{kl}|/s)$ as
well as the SL contributions
proportional to $\ls$ are denoted as the symmetric-electroweak part of
the corrections.  The IR singularities are regularized by an
infinitesimal photon mass $\la$, and owing to the mass hierarchy
\beq
\MH,\Mt,\MW,\MZ \gg m_{f\ne\Pt} \gg  \lambda,
\eeq
all logarithms of electromagnetic origin $\lWla$ and $\lWf$ involving
the photon mass or light charged fermion masses are large and have to be
taken into account, whereas the logarithms $\lWZ$, $\ltW$, and $\lHW$
are neglected. Furthermore, in the limit \refeq{Sudaklim}, the pure
angular-dependent contributions $\log{(r_{kl}/s)}$ and
$\log^2{(r_{kl}/s)}$ can be neglected.
 
The lowest-order matrix element for \refeq{process} is denoted by
\begin{equation}
\M_0^{i_1 \ldots i_n}(p_1,\ldots, p_n).
\end{equation}
In LA the corrections assume the form  
\beq \label{LAfactorization} 
\delta \M^{i_1 \ldots i_n}(p_1,\ldots, p_n)= 
\M_0^{i'_1 \ldots i'_n}(p_1, \ldots, p_n)\delta_{i'_1i_1 \ldots i'_ni_n},
\end{equation}
\ie they factorize as a matrix, 
and are split into various contributions according to their origin:
\beq
\delta=\de^{\SC}+\de^{\SS}+\de^{\cc}+\de^\pre.
\eeq
The leading and subleading soft--collinear logarithms are denoted by
$\de^\SC$ and $\de^\SS$, respectively, the collinear logarithms by
$\de^\cc$, and the logarithms resulting from parameter
renormalization, which can be determined by the running of the
couplings, by $\de^\pre$. 

\section{Soft--collinear contributions}
\label{se:soft-coll}

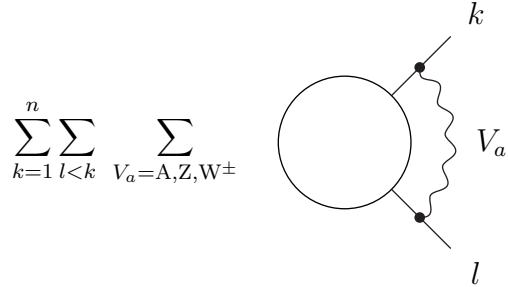
\begin{figure}[h] 
\begin{center}
\begin{picture}(140,110)
\Line(70,50)(110,90)
\Line(70,50)(110,10)
\PhotonArc(70,50)(40,-45,45){2}{5}
\GCirc(70,50){25}{1}
\Text(30,50)[r]{${\displaystyle \sum\limits_{k=1}^n\sum\limits_{l<k}\,\,\sum\limits_{\GB_a=\PA,\PZ,\PW^\pm}}$}
\Text(120,50)[l]{$\GB_a$}
\Text(120,95)[b]{$k$}
\Text(120,5)[t]{$l$}
\Vertex(98.3,78.3){2}
\Vertex(98.3,21.7){2}
\end{picture}
\end{center}
\caption{Feynman diagrams leading to DL corrections}
\label{SCdiag}
\end{figure}

The DL corrections originate from loop diagrams where virtual gauge
bosons $\GB_a=\PA,\PZ,\PW^\pm$ 
are exchanged between pairs of external
legs (\reffi{SCdiag}).
The double logarithms arise from the integration region where the
gauge-boson momenta are soft and collinear to one of the external
legs. As in QED, they can be evaluated using the eikonal
approximation, where in the numerator of the loop integral the
gauge-boson momentum is set to zero and all mass terms are neglected.
In this approximation the one-loop corrections give
\beq \label{eikonalappA}
\delta \M^{i_1 \ldots i_n}=\sum_{k=1}^n\sum_{l<
  k}\sum_{\GB_a=A,Z,W^\pm}\int \frac{\rd^4q}{(2\pi)^4} \frac{-4ie^2
  p_kp_lI^{\GB_a}_{i'_k i_k}(k) I^{\bar{\GB}_a}_{i'_l i_l}(l) \M_0^{i_1 \ldots i'_k
    \ldots i'_l \ldots i_n}}
{(q^2-M_{\GB_a}^2)[(p_k+q)^2-m_{k'}^2][(p_l-q)^2-m_{l'}^2]},
\eeq
and in LA, using the high-energy expansion of the scalar three-point
function \cite{Denn3}, one obtains
\beq \label{eikonalapp}
\delta \M^{i_1 \ldots i_n}
=\frac{1}{2}\sum_{k=1}^n\sum_{l\neq k}\sum_{\GB_a=A,Z,W^\pm} I^{\GB_a}_{i'_k i_k}(k) I^{\bar{\GB}_a}_{i'_l i_l}(l) \M_0^{i_1 \ldots i'_k \ldots i'_l \ldots i_n}[\LrMa-\de_{\GB_aA}\Lkla].
\eeq
The DL term containing the invariant $r_{kl}$ depends on the angle
between the momenta $p_k$ and $p_l$. Writing
\beq \label{angsplit}
\LrM=\LsM+2\lsM\lrs+\Lrs,
\eeq
the angular-dependent part is isolated in logarithms of $r_{kl}/s$,
and gives a subleading soft--collinear ($\SS$) contribution of order
$\ls\log(|r_{kl}|/s)$, whereas terms $\Lrs$ can be neglected in LA.
The remaining part, together with the 
additional contributions from photon
loops in \refeq{eikonalapp}, gives the leading soft--collinear ($\SC$)
contribution and is angular-independent.  The eikonal approximation
\refeq{eikonalappA} applies to chiral fermions, Higgs bosons, and
transverse gauge bosons, and depends on their gauge couplings
$I^{\GB_a}(k)$.

Owing to the longitudinal polarization vectors
\refeq{longplovec} which grow with energy, matrix elements involving longitudinal gauge bosons have to be treated with the equivalence theorem, \ie they have to be expressed by matrix elements involving the corresponding Goldstone bosons. 
A detailed description of the equivalence theorem is given in \refse{loggaugebos}. As explained there, the equivalence theorem for Born matrix elements \refeq{eq:borneet} receives no DL one-loop corrections. Therefore, the soft-collinear  corrections for external longitudinal gauge bosons can be obtained using the simple relations
\beqar \label{DLeqtheor}
\de^{\mathrm{DL}}\M^{\ldots \PW^\pm_\rL \ldots} &=&\de^{\mathrm{DL}}\M^{\ldots \phi^\pm \ldots},\nl
\de^{\mathrm{DL}}\M^{\ldots \PZ_\rL \ldots} &=&\ri\de^{\mathrm{DL}}\M^{\ldots \chi \ldots},
\eeqar
from the corrections \refeq{eikonalapp} for external Goldstone bosons.

\subsection*{Leading soft--collinear contributions}
The invariance of the S matrix with respect to global
$\SUtwo\times\Uone$ transformations implies 
\beq\label{eq:gi}
0= \delta_{\GB_a} 
\M^{i_1 \ldots i_n}=\ri e \sum_{k} I^{\GB_a}_{i'_k i_k}(k) \M^{i_1 \ldots i'_k \ldots i_n}.
\eeq
For external Goldstone fields extra contributions proportional to the
Higgs vacuum expectation value appear, which are, however, irrelevant
in the high-energy limit.  Using \refeq{eq:gi}, the $\SC$ logarithms
in \refeq{eikonalapp} can be written as a single sum over external legs,
\beq \label{SCsum}
\de^{\SC} \M^{i_1 \ldots i_n} =\sum_{k=1}^n \delta^\SC_{i'_ki_k}(k)
\M_0^{i_1 \ldots i'_k\ldots i_n}.
\end{equation} 
After evaluating the sum over  A, Z, and W in \refeq{eikonalapp}, the
correction factors read
\beq \label{deSC} 
\de^\SC_{i'_ki_k}(k)=- \frac{1}{2}\left[ C^{\ew}_{i'_ki_k}(k)\Ls -2(I^Z(k))_{i'_ki_k}^2 \log{\frac{\MZ^2}{\MW^2}}\, \ls+\de_{i'_ki_k} Q_k^2\Lemk \right].
\eeq
The first term represents the DL symmetric-electroweak part and is
proportional to the electroweak Casimir operator $\cew$ defined in
\refeq{CasimirEW}. This is always diagonal in the $\SUtwo$ indices,
except for external transverse
neutral gauge bosons in the physical basis
\refeq{adjointCasimir2}, where it gives rise to mixing between
amplitudes involving photons and Z bosons. The second term originates
from Z-boson loops, owing to the difference between $\MW$ and $\MZ$,
and
\beq
\Lemk:= 2\ls\log{\left(\frac{\MW^2}{\la^2} \right)}+\LWla-\Lkla
\eeq
contains all logarithms of pure electromagnetic origin.  
The $\SC$ corrections for external longitudinal gauge bosons are directly obtained from \refeq{deSC} by using the quantum numbers of the corresponding Goldstone bosons.
Formula \refeq{deSC} is in agreement with \citeres{Ku1,Fa00}. In
\citere{CC1} the logarithm $\Lkla$ that depends on the mass of the external state is missing.

\subsection*{Subleading soft--collinear contributions}
The contribution of the second term of \refeq{angsplit} to
\refeq{eikonalapp} remains a sum over pairs of external legs,
\beq \label{SScorr}
\de^\SS \M^{i_1 \ldots i_n} =\sum_{k=1}^n
\sum_{l<k}\sum_{\GB_a=A,Z,W^\pm}\delta^{\GB_a,\SS}_{i'_ki_k i'_li_l}(k,l)
\M_0^{i_1\ldots i'_k\ldots i'_l\ldots i_n},
\eeq
with angular-dependent terms. The exchange of soft, neutral gauge
bosons contributes with
\beqar \label{subdl1} 
\de^{A,\SS}_{i'_ki_k i'_li_l}(k,l)&=&
2 \left[\ls+\lWla\right]\lrs I_{i'_ki_k}^A(k)I_{i'_li_l}^A(l),\nl
\delta^{Z,\SS}_{i'_ki_k i'_li_l}(k,l)&=&
2\ls \lrs I_{i'_ki_k}^Z(k)I_{i'_li_l}^Z(l),
\eeqar
and, except for $I^Z$ in the neutral scalar sector $H,\chi$ (see
\refapp{app:representations}),  
the couplings $I^\NB$ are diagonal matrices.  
The exchange of charged gauge bosons yields
\beq \label{subdl2} 
\delta^{W^\pm,\SS}_{i'_ki_k i'_li_l} (k,l)=2\ls\lrs I_{i'_ki_k}^\pm(k)I_{i'_li_l}^{\mp}(l),
\eeq
and owing to the non-diagonal matrices $I^\pm(k)$ [\cf
\refeq{ferpmcoup}, \refeq{scapmcoup} and \refeq{gaupmcoup}], 
contributions of $\SUtwo$-transformed Born matrix elements appear on
the left-hand side of \refeq{SScorr}. 
In general,
these transformed Born matrix elements are not related to the 
original Born matrix element and have to be evaluated explicitly.

The $\SS$ corrections for external longitudinal gauge bosons are obtained from \refeq{SScorr} with the equivalence theorem \refeq{DLeqtheor}
, \ie the couplings and the Born matrix elements for Goldstone bosons\footnote{Note that for Goldstone bosons $\chi$, the equivalence theorem as well as the couplings \refeq{scapmcoupB} and \refeq{ZHcoup} contain the imaginary constant $\ri$.}  have to be used on the right-hand side of \refeq{SScorr}. 

The application of the above formulas is illustrated in \refse{se:applicat} for the case of 4-particle processes, where  
owing to $r_{12}=r_{34}$,
$r_{13}=r_{24}$ and  $r_{14}=r_{23}$, 
\refeq{SScorr} reduces to
\beqar  \label{4fsubdl} 
\de^\SS \M^{i_1i_2i_3i_4} &=&\sum_{\GB_a=A,Z,W^\pm}2 \left[\ls+\lWa\right]\times\\
&&\left\{\log{\frac{|r_{12}|}{s}}\left[
I^{\GB_a}_{i_1'i_1}(1)
I^{\bar{\GB}_a}_{i_2'i_2}(2) 
\M_0^{i'_1i'_2i_3i_4}
+I^{\GB_a}_{i_3'i_3}(3)I^{\bar{\GB}_a}_{i'_4i_4}(4) 
\M_0^{i_1i_2i'_3i'_4}
\right]\right.\nl
&&\left.{}+\log{\frac{|r_{13}|}{s}}\left[
I^{\GB_a}_{i_1'i_1}(1)I^{\bar{\GB}_a}_{i_3'i_3}(3) 
\M_0^{i'_1i_2i'_3i_4}
+I^{\GB_a}_{i_2'i_2}(2)I^{\bar{\GB}_a}_{i_4'i_4}(4) 
\M_0^{i_1i'_2i_3i'_4}
\right]\right.\nl
&&\left.{}+\log{\frac{|r_{14}|}{s}}\left[
I^{\GB_a}_{i_1'i_1}(1)I^{\bar{\GB}_a}_{i_4'i_4}(4) 
\M_0^{i'_1i_2i_3i'_4}
+I^{\GB_a}_{i_2'i_2}(2)I^{\bar{\GB}_a}_{i_3'i_3}(3) 
\M_0^{i_1i'_2i'_3i_4}
\right]\right\},\nonumber
\eeqar
and the logarithm with $r_{kl}=s$ vanishes.
Note that this formula applies to $4\rightarrow 0$ processes, where
all particles or antiparticles and their momenta are
incoming. Predictions for $2\rightarrow 2$ processes  are obtained by
substituting outgoing particles (antiparticles) by the corresponding incoming
antiparticles (particles), as explained in \refse{se:defcon}.

\section{Collinear and soft single logarithms}
\label{se:soft-or-coll}
In this section we consider the SL corrections 
originating from field renormalization and from mass-singular loop diagrams.  
The PR contributions associated with the renormalization of the electric
charge, the weak mixing angle, and mass ratios are presented in 
\refse{se:ren}.  As explained in the introduction, in our approach to SL 
corrections we set the regularization scale $\mu^2=s$ so that only 
mass-singular logarithms $\log{(\mu^2/M^2)}$ or $\log{(s/M^2)}$ are large. 

On one hand the FRCs give the well-known
factors $\de Z_{\varphi}/2$ for each external leg, containing
collinear as well as soft SL contributions.
On the other hand,
mass-singular logarithms arise from the collinear limit of loop
diagrams where an external line splits into two internal lines
\cite{KLN}, one of these internal lines being a virtual gauge boson
$\PA,\PZ$ or $\PW$.  
If the two internal lines involve only fermions and scalars no mass
singular terms emerge.
The mass-singular diagrams are evaluated in the limit of
collinear gauge-boson emission using Ward identities \cite{DennPozz2},
and after subtraction of the contributions already contained in the
FRCs and in the soft--collinear corrections, 
we find factorization into collinear factors $\de^\coll$ times Born
matrix elements,
\beqar
\lefteqn{
\sum_{\GB_a=A,Z,W^\pm} \left\{
\vcenter{\hbox{\begin{picture}(100,80)(0,-40)
\Line(65,0)(85,0)
\Line(65,0)(39.1,14.1)
\Photon(65,0)(39.1,-14.1){2}{3}
\Vertex(65,0){2}
\GCirc(25,0){20}{1}
\Text(50,-18)[l]{$\GB_a$}
\Text(85,5)[rb]{$k$}
\end{picture}}}
-
\vcenter{\hbox{\begin{picture}(100,80)(0,-40)
\Line(25,0)(94,0)
\PhotonArc(67,0)(12,180,0){1.5}{4}
\Vertex(55,0){2}
\Vertex(79,0){2}
\GCirc(25,0){20}{1}
\Text(67,-18)[t]{$\GB_a$}
\Text(94,5)[rb]{$k$}
\end{picture}}}\right.
}\qquad\nl
&& \left.\left. - \sum_{l\neq k}\left[
\vcenter{\hbox{\begin{picture}(100,90)(0,-45)
\Line(25,0)(65,40)
\Line(25,0)(65,-40)
\PhotonArc(25,0)(30,-45,45){2}{4}
\Vertex(46.2,21.2){2}
\Vertex(46.2,-21.2){2}
\GCirc(25,0){20}{1}
\Text(60,0)[l]{$\GB_a$}
\Text(70,35)[lb]{$k$}
\Text(70,-35)[lt]{$l$}
\end{picture}}}
\right]_{\mathrm{eik.~appr.}}\right\}\right|_{\mathrm{coll.}}
=\de^\coll(k)
\vcenter{\hbox{\begin{picture}(80,80)(0,-40)
\Line(25,0)(70,0)
\GCirc(25,0){20}{1}
\Text(70,5)[rb]{$k$}
\end{picture}}}.
\eeqar
Then, the complete SL contributions 
originating from soft or collinear regions
can be written as a sum over the external legs,
\beq\label{subllogfact}
\de^{\cc} \M^{i_1 \ldots i_n} =\sum_{k=1}^n \delta^\cc_{i'_ki_k}(k)
\M_0^{i_1 \ldots i'_k \ldots i_n}
\eeq
with 
\beq\label{subllogfact2}
 \delta^\cc_{i'_ki_k}(k)=\left.\delta^\coll_{i'_ki_k}(k)+\frac{1}{2}\delta Z^\varphi_{i'_k i_k}\right|_{\mu^2=s}.
\eeq
The collinear factors $\de^\coll(k)$ and the corrections
$\de^\cc(k)$ depend on the quantum numbers of the external fields
$\varphi_{i_k}$.
 In the following we give the results for chiral fermions,
transverse charged gauge bosons $\PW_\rT$, transverse neutral gauge
bosons $\PA_\rT,\PZ_\rT$, longitudinal gauge bosons $\PW_\rL,\PZ_\rL$,
and Higgs bosons. 
We use the conventions of \citere{FortPhys} for the Feynman rules, the 
self-energies, and the renormalization constants.

\subsection*{Chiral fermions}
In LA the FRCs for fermions $f^\kappa_\si$ with chirality $\kappa=\rR,\rL$ and isospin indices $\si=\pm$ are given by
\beqar
\delta Z^\kappa_{f_\si f_{\si'}} &=&\de_{\si\si'}\left\{-\left[\cew_{f^\kappa}+\frac{1}{4\sw^2}\left((1+\delta_{\kappa \rR})\frac{m^2_{f_\si}}{\MW^2}+\delta_{\kappa \rL}\frac{m_{f_{-\si}}^2}{\MW^2}\right)\right] \lu \right.\nl&&
\left.{}\hspace{0.8cm}+ Q_{f_\si}^2\left[2\lWla -3\lWfsi\right]\right\},
\eeqar
where the contribution of a non-trivial quark-mixing matrix is not
considered. The FRCs depend on the chirality of the fermions, and
contain Yukawa terms proportional to the masses of the fermion $f_\si$ and
of its isospin partner $f_{-\si}$. While these are negligible for leptons
and light quarks, they give large contributions for $f^\kappa_\si=\Pt^\rR$, 
  $\Pt^\rL$, and $\Pb^\rL$, where one of the masses is $\Mt$. 

From the mass-singular loop diagrams
we obtain the factor \cite{DennPozz2}
\beq \label{eq:collf}
\de^{\coll}_{f_\si f_{\si'}}(f^\kappa)=
\de_{\si\si'}\left[2\cew_{f^\kappa}\lu +2 Q_{f_\si}^2\lWfsi\right],
\eeq
and the complete contribution \refeq{subllogfact2} reads
\beq \label{deccfer}
\de^{\cc}_{f_\si f_{\si'}}(f^\kappa)=\de_{\si\si'}\left\{\left[\frac{3}{2} \cew_{f^\kappa} -\frac{1}{8\sw^2}\left((1+\delta_{\kappa \rR})\frac{m_{f_\si}^2}{\MW^2}+\delta_{\kappa \rL}\frac{m_{f_{-\si}}^2}{\MW^2}\right)\right]\ls+Q_{f_\si}^2\lemfsi\right\},
\eeq
where the pure electromagnetic logarithms
\beq \label{lemf}
\lemf:=\frac{1}{2}\lWf +\lWla
\eeq
originate from the photonic loops as a result of the gap between the
electromagnetic and weak scales.
The symmetric-electroweak part of
\refeq{deccfer}, \ie the term proportional to $\ls$,  
agrees with \citeres{Ku2,Me1} up to the Yukawa
contributions, and the electromagnetic part \refeq{lemf} agrees with
\citere{Me1}. 
\subsection*{Transverse charged gauge bosons W}
The FRC of $\PWpm$ bosons in LA reads
\beq \label{deltaWwf}
\delta Z_{WW}=
-\left.\frac{\partial\Sigma^{WW}_{\rT}(k^2)}{\partial k^2}\right|_{k^2=\MW^2} 
=\left[\bew_{W}-2\cew_{W} \right]\lu +2Q_\PW^2\lWla,
\eeq
where $\bew_{W}$ is the coefficient of the $\beta$-function 
defined in
\refeq{betafunction} and contains the sum over gauge-boson, scalar,
and fermion loops, whereas $\cew_{W}$ 
is the 
 eigenvalue of the electroweak Casimir operator
in the
adjoint representation \refeq{gaugeeigenvalues}.

Combining the collinear factor \cite{DennPozz2}
\beq 
\de^{\coll}_{W^\si W^{\si'}}(\GB_\rT) =\de_{\si\si'}\cew_{W}\lu
\eeq
with the FRC $\delta Z_{W^\si W^{\si'}}=\de_{\si\si'}\delta Z_{WW}$,
results in
\beq \label{deccWT}
\delta^\cc_{W^\si W^{\si'}}(\GB_{\rT})=\de_{\si\si'}\left[\frac{1}{2}\bew_{W}\ls +Q_\PW^2\lemW\right].
\eeq
This result agrees with the revised version of \citere{Me1}. 
 
\subsection*{Transverse neutral gauge bosons A,Z}
\newcommand{\antikro}{E}
\newcommand{\kroAA}{\de_{\NB A}\de_{\NB'A}}
\newcommand{\kroZZ}{\de_{\NB Z}\de_{\NB'Z}}
The physical neutral gauge-boson fields $\NB=\PA,\PZ$ are renormalized
by a non-symmetric matrix $\de Z$, \ie
\beq \label{phWFRCngb}
\NB\to \NB+\de \NB,\qquad \delta \NB = \frac{1}{2}\delta Z_{\NB\NB'} \NB' = \frac{1}{2}\left[\delta Z^\asymm_{\NB\NB'}+\delta Z^\symm_{\NB\NB'}\right] \NB'.
\eeq
The matrix $\de Z$ has been split into antisymmetric and symmetric
parts in order to facilitate the comparison with the corresponding FRC
$\de\tilde Z$ for the symmetric components $\sNB=B,W^3$,
\beq \label{sWFRCngb}
\de \sNB=\frac{1}{2}\de \tilde{Z}_{\sNB\sNB'}\sNB'.
\eeq
In the following, we give $\de Z_{\NB\NB'}$ as obtained in LA in the
on-shell scheme \cite{FortPhys} and compare it with $\de
\tilde{Z}_{\sNB\sNB'}$ using the matrix relation
\beq \label{unbrokenWFREN} 
\delta Z=2 \delta U(\thw) U^{-1}(\thw)+ U(\thw)\delta \tilde{Z} U^{-1}(\thw),
\eeq
resulting from the renormalization of the Weinberg rotation $U(\thw)$
defined in \refeq{weinbergrotation}.

The results for symmetric and antisymmetric parts are expressed in
terms of the coefficients of the $\beta$-function 
defined in  \refeq{vvbetafunction}:
\begin{itemize}
\item For the antisymmetric part the diagonal components vanish,
  whereas the non-diagonal ones are
\beq  \label{deltaZa0}
\delta Z^\asymm_{AZ}=-\delta Z^\asymm_{ZA}= -\frac{\Sigma^{AZ}_{\rT}(\MZ^2)+\Sigma^{AZ}_{\rT}(0)}{\MZ^2},
\eeq 
and in LA we find
\beq \label{deltaZa}
\delta Z^\asymm_{AZ}= \bew_{AZ} \lu.
\eeq
This part is related to the renormalization of the Weinberg angle
\refeq{weinbergrenorm},
and in LA it corresponds to the first term in \refeq{unbrokenWFREN}
\beq \label{WRMAT}
2 \left[\delta U(\thw) U^{-1}(\thw)\right]_{\NB\NB'}=\frac{\cw}{\sw}\frac{\delta \cw^2}{\cw^2}\antikro_{\NB\NB'}=\bew_{AZ}\lu \antikro_{\NB\NB'}
,\qquad \antikro:=\left(\begin{array}{c@{\;}c} 0& 1  \\ -1 & 0
  \end{array}\right).
\eeq
\item The symmetric part has components 
\beq
\delta Z^\symm_{\NB\NB'}=
-\left.\frac{\partial\Sigma^{\NB\NB'}_{\rT}(k^2)}{\partial k^2}\right|_{k^2=M_\NB^2}, 
\qquad \delta Z^\symm_{AZ}=\delta Z^\symm_{ZA}=-\frac{\Sigma^{AZ}_{\rT}(\MZ^2)-\Sigma^{AZ}_{\rT}(0)}{\MZ^2},
\eeq
and in LA it reads
\beq \label{deltaZs}
\delta{Z}^\symm_{\NB\NB'}= \left[\bew_{\NB\NB'}-2\cew_{\NB\NB'}\right]\lu
+\kroAA \de Z^\elm_{AA} .
\eeq
The $AA$ component receives a pure electromagnetic contribution
associated with the light-fermion loops,
\beq
\de Z^\elm_{AA}=-\frac{4}{3}\sum_{f,i,\si\neq t} \NCf Q^2_{f_\si}\lWfsii,
\eeq
where the sum runs over the generations $i=1,2,3$ of leptons and
quarks $f=l,q$ with isospin $\si$, omitting the top-quark
contribution. 

Apart from these pure electromagnetic logarithms, \refeq{deltaZs} contains
the same combination of $\bew$ and $\cew$ as \refeq{deltaWwf}. This
part corresponds to the second term in \refeq{unbrokenWFREN}
originating from the renormalization of the symmetric fields,
\beq
\delta \tilde{Z}_{\sNB\sNB'}=\delta_{\sNB\sNB'}\left[ \besw_{\sNB}-2\csew_{\sNB} \right]\lu, 
\eeq
which is diagonal, because the $\Uone$ and $\SUtwo$ components do not
mix in the unbroken theory. 
\end{itemize}

The SL contributions \refeq{deltaZa} and \refeq{deltaZs} have to
be combined with the collinear factor, for which  we obtain  \cite{DennPozz2}
\beq
\de^{\coll}_{\NB\NB'}(\GB_\rT) =\cew_{\NB\NB'}\lu.
\eeq
Then, the complete correction is given by
\beq \label{deccVVT} 
\de^\cc_{\NB'\NB}(\GB_{\rT})= \frac{1}{2}\left[\antikro_{\NB'\NB}\bew_{AZ} +\bew_{\NB'\NB} \right]\ls +\frac{1}{2}\kroAA \de Z^\elm_{AA}. 
\eeq
Note that owing to the antisymmetric contribution ($E_{AZ}=-E_{ZA}=1$)  the non-diagonal components read
\beq
\de^\cc_{AZ}(\GB_{\rT})=\bew_{AZ}\ls,\qquad
\de^\cc_{ZA}(\GB_{\rT})=0,
\eeq
\ie the correction factor for external photons does not involve mixing
with $\PZ$ bosons.  
This is a consequence of the on shell renormalization condition \refeq{deltaZa0}.   
The symmetric  part of \refeq{deccVVT} agrees
with the revised version of \citere{Me1}. 

\subsection*{Longitudinally polarized gauge bosons}\label{loggaugebos}
Our approach is not directly applicable to
the calculation of the effective collinear factor \refeq{subllogfact}
for longitudinal gauge bosons, because the amputated Green functions
involving gauge bosons are contracted with longitudinal polarization
vectors,
\beq \label{longplovec}
\epsilon_\rL^\mu(p)=\frac{p^\mu}{M}+\O\left(\frac{M}{p^0}\right),
\eeq 
containing a mass term in the denominator so that in this case
contributions of the order of the gauge-boson mass cannot be
neglected.  This problem can be circumvented by means of the
Goldstone-boson equivalence theorem, expressing the Green functions
involving longitudinal gauge bosons by Green functions with the
corresponding Goldstone bosons. The equivalence theorem for bare
amputated Green function reads (we denote bare quantities by an index 0)
\beqar \label{eq:bareet} 
p^\mu\langle W_{0,\mu}(p)\ldots\rangle&=&M_{0,W}(1+\de C_{W_0})\langle \phi_0(p) \ldots\rangle,\nl
p^\mu\langle Z_{0,\mu}(p)\ldots\rangle &=&\ri M_{0,Z}(1+\de C_{Z_0})\langle \chi_0(p) \ldots\rangle,
\eeqar
where the dots represent arbitrary fields. 
In Born approximation, this gives the well-known relations
\beqar \label{eq:borneet} 
\M_0^{\ldots \PW^\pm_\rL \ldots} &=&\M_0^{\ldots \phi^\pm \ldots},\nl
\M_0^{\ldots \PZ_\rL \ldots} &=&\ri\M_0^{\ldots \chi \ldots},
\eeqar
between matrix elements.
Note however, that besides the lowest-order contribution,
\refeq{eq:bareet} contains
non-trivial higher-order corrections $\de C_{W_0}$, $\de C_{Z_0}$
owing to the mixing between gauge bosons and Goldstone bosons
\cite{etcorr}. In one-loop approximation these corrections can be
expressed in terms of bare self-energies involving Goldstone bosons
and longitudinal gauge bosons evaluated at the mass of the gauge
bosons,
\beqar\label{eq:bareetcorr}
\de C_{W_0} &=&  - \frac{\Si^{WW}_{\rL}(\MW^2)+\MW \Si^{W\phi}(\MW^2)}{\MW^2},\nl 
\de C_{Z_0} &=&  - \frac{\Si^{ZZ}_{\rL}(\MZ^2)-\ri\MZ \Si^{Z\chi}(\MZ^2)}{\MZ^2}.
\eeqar
Since neither $\de C$ nor the counterterms involve double logarithms,
the equivalence theorem can be applied to the DL corrections in the
naive way, \ie without higher-order corrections $\de C_{W_0}$, $\de
C_{Z_0}$.

The renormalization of \refeq{eq:bareet} leads to extra mass and
field-renormalization counterterms. Especially, the renormalization in
the neutral sector involves mixing effects, but as expected, the
physical longitudinal Z boson does not mix with the photon.
Keeping the unphysical scalar fields unrenormalized, and absorbing correction
factors and counterterms into new renormalized correction factors $\de
C_{\phi}$, $\de C_{\chi}$, we can write
\beqar \label{eq:renet} 
p^\mu\langle W_{\mu}(p)\ldots\rangle&=&\MW(1+\de C_{\phi})\langle \phi_0(p) \ldots\rangle,\nl
p^\mu\langle Z_{\mu}(p)\ldots\rangle&=&\ri\MZ(1+\de C_{\chi})\langle\chi_0(p) \ldots\rangle
\eeqar
with 
\beqar \label{eq:renetcorrA} 
\de C_\phi&=&\de C_{W_0}+\frac{\de \MW}{\MW}+\frac{1}{2}\de Z_{WW},\nl
\de C_\chi&=&\de C_{Z_0}+\frac{\de \MZ}{\MZ}+\frac{1}{2}\de Z_{ZZ}.
\eeqar  
In LA we find
\beqar \label{eq:renetcorr} 
\de C_\phi&=&\left[\cew_\Phi-\frac{\NCt}{4\sw^2}\frac{\Mt^2}{\MW^2}\right]\lu
+Q_\PW^2\lWla,\nl
\de C_\chi&=&\left[\cew_\Phi-\frac{\NCt}{4\sw^2}\frac{\Mt^2}{\MW^2}\right]\lu.
\eeqar
The result is written in terms of the eigenvalue of $\cew$ for the
scalar doublet $\Phi$ and contains large $\Mt$-dependent contributions
originating from the mass counterterms \refeq{massCT}, which are 
proportional to the colour factor $\NCt=3$.  With \refeq{eq:renet} and
with the collinear factor for Goldstone bosons $S=\phi^\pm,\chi$
\cite{DennPozz2},
\beq \label{eq:collsGB}
\de^{\coll}_{SS'}(\Phi)=\de_{SS'}\cew_\Phi\lu,
\eeq
the complete collinear corrections  \refeq{subllogfact} for longitudinal gauge bosons are obtained by means of amplitudes involving Goldstone bosons,
\beqar
\de^{\cc} \M^{\ldots \PW^\pm_\rL \ldots} &=&\left[\de^{\coll}_{\phi^\pm\phi^\pm}(\Phi)+\de C_\phi\right] \M_0^{\ldots \phi^\pm \ldots}=\de^{\cc}_{\phi^\pm\phi^\pm}(\Phi)\, \M_0^{\ldots \PW^\pm_\rL \ldots},\nl
\de^{\cc} \M^{\ldots \PZ_\rL \ldots} &=&\ri\left[\de^{\coll}_{\chi\chi}(\Phi)+\de C_\chi\right]  \M_0^{\ldots \chi \ldots}=\de^{\cc}_{\chi\chi}(\Phi)\, \M_0^{\ldots \PZ_\rL \ldots}
\eeqar
with
\beqar \label{longeq:coll} 
\de^\cc_{\phi^\pm\phi^\pm}(\Phi)&=&  \left[2\cew_\Phi-\frac{\NCt}{4\sw^2}\frac{\Mt^2}{\MW^2}\right]\ls   +Q_\PW^2\lemW ,\nl
\de^\cc_{\chi\chi}(\Phi)&=& \left[2\cew_\Phi-\frac{\NCt}{4\sw^2}\frac{\Mt^2}{\MW^2}\right]\ls.
\eeqar 
Note that, up to pure electromagnetic terms, the correction factors
\refeq{eq:renetcorrA} correspond to FRCs for Goldstone bosons. In fact,
in LA $\de C_\chi=\de Z_\chi/2$ and $\de
C_\phi=\de Z_\phi/2 + Q_\PW^2\lWla$.
\subsection*{Higgs bosons}
In LA the FRC for Higgs bosons reads
\begin{equation}
\delta Z_{H} = \left[2\cew_\Phi-\frac{\NCt}{2\sw^2}\frac{\Mt^2}{\MW^2}\right]\lu
\end{equation}
with a large Yukawa contribution coming from the top-quark loop. 
For the collinear factor we find \cite{DennPozz2}
\beq \label{eq:collsH}
\de^{\coll}_{HH}(\Phi)=\cew_\Phi\lu,
\eeq
and the complete correction is
\beq
\de^{\cc}_{HH}(\Phi)= \left[2\cew_\Phi-\frac{\NCt}{4\sw^2}\frac{\Mt^2}{\MW^2}\right]\ls.
\eeq
Note that up to pure electromagnetic contributions, longitudinal gauge
bosons and Higgs bosons receive the same collinear SL corrections.

\section{Logarithms connected to parameter renormalization}
\label{se:ren}
\newcommand{\eff}{\mathrm{eff}}
\newcommand{\gt}{g_{\Pt}}
\newcommand{\gH}{\la}
\newcommand{\rt}{h_{\Pt}}
\newcommand{\rH}{h_{\PH}}

Finally, there are logarithms related to UV divergences. These
logarithms originate from the renormalization of the dimensionless
parameters, \ie the electric charge
$e$, the weak mixing angle $\cw$, and the mass ratios
\beq
\rt =\frac{\Mt}{\MW}, \qquad  \rH =\frac{\MH^2}{\MW^2},
\eeq
and are obtained from the Born matrix element $\M_0=\M_0(e,\cw,\rt,\rH)$
in the high-energy limit by
\beq
\de^\pre \M = \left. \frac{\de\M_0}{\de e}\de e 
+ \frac{\de\M_0}{\de\cw}\de\cw 
+ \frac{\de\M_0}{\de\rt}\de\rt 
+ \frac{\de\M_0}{\de\rH}\de\rH^{\eff} 
\, \right|_{\mu^2=s}
.
\eeq
The mass ratios $\rt$ and $\rH$ are related to the top-quark Yukawa coupling
and to the scalar self coupling, respectively. They appear only in
processes where these couplings enter. The renormalization of the
masses in the propagators or in couplings with mass dimension yields
only mass-suppressed contributions which are irrelevant in the
high-energy limit in amplitudes that are not mass-suppressed.

The logarithms connected to parameter renormalization can simply be
obtained by the replacements $e\to e+\de e$, $\cw\to\cw+\de\cw$,
$\sw\to\sw+\de\sw$, $\rt\to\rt+\de\rt$ and $\rH\to\rH+\de\rH^{\eff}$ in the
lowest-order matrix elements in the high-energy limit. In the case of
processes with longitudinal gauge bosons, these substitutions must
be performed in the matrix elements resulting from the equivalence
theorem.
 
\subsection*{Mixing-angle renormalization}
In the on-shell scheme, the renormalization of the weak mixing angle
\refeq{mixingangle} is given by
\beq  \label{Weinren}
\frac{\delta \cw^2}{\cw^2}=\frac{\de\MW^2}{\MW^2} -\frac{\de\MZ^2}{\MZ^2}=\frac{\Si_{T}^{W}(\MW^2)}{\MW^2} -\frac{\Si_{T}^Z(\MZ^2)}{\MZ^2}.
\eeq
After tadpole renormalization, \ie omitting the tadpole diagrams,
the mass counterterms give
\beqar \label{massCT}
\frac{\de\MW^2}{\MW^2} &=&-\left[\bew_{W}-4\cew_\Phi\right]\lu -\frac{\NCt}{2\sw^2}\frac{\Mt^2}{\MW^2}\lu, \nl
\frac{\de\MZ^2}{\MZ^2} &=&-\left[\bew_{ZZ} -4\cew_\Phi\right]\lu -\frac{\NCt}{2\sw^2}\frac{\Mt^2}{\MW^2}\lu,
\eeqar
and contain large $({\Mt^2}/{\MW^2})\lu$ terms.
However, these terms cancel in \refeq{Weinren}, and using 
\beq
\bew_{AZ}=\frac{\cw}{\sw}(\bew_{ZZ}-\bew_{W}),
\eeq
which follows from \refeq{betarelations},
we can express the mixing-angle counterterm by the $AZ$ component of
the $\beta$-function:
\beq \label{weinbergrenorm}
\frac{\delta \cw^2}{\cw^2}=\frac{\sw}{\cw}\bew_{AZ}\lu.
\eeq

\subsection*{Charge renormalization}
In the on-shell scheme, the coupling-constant counterterms are related to the FRCs by Ward identities. For the electric charge counterterm we have
\beqar \label{chargerenorm}
\de Z_e&=& -\frac{1}{2}\left[\de Z_{AA}+\frac{\sw}{\cw}\de
  Z_{ZA}\right]
=\left.\frac{1}{2}\frac{\partial \Si_{T}^{AA}(k^2)}{\partial k^2}\right|_{k^2=0}
-\frac{\sw}{\cw}\frac{\Si_{T}^{AZ}(0)}{\MZ^2}\nl
&=&-\frac{1}{2}\bew_{AA}\lu+\de Z^{\elm}_e,
\eeqar
where the pure electromagnetic part
\beq
\de Z^{\elm}_e= -\frac{1}{2}\de Z^\elm_{AA}=\frac{2}{3}\sum_{f,i,\si \neq t} \NCf Q_{f_\si}^2 \lWfsii,
\eeq
is related to the running of the electromagnetic coupling constant
from zero momentum transfer to the electroweak scale,
\beq \label{runningal}
\Delta\al(\MW^2)=2\de Z^\elm_e.
\eeq
The counterterms to the $\Uone$ and $\SUtwo$ gauge couplings 
\beq \label{gaugecouplings}
g_1=\frac{e}{\cw},\qquad g_2=\frac{e}{\sw}
\eeq
can be written as 
\beqar \label{gCTs}
\frac{\de g_1}{g_1}&=&\de Z_e -\frac{1}{2}\frac{\delta \cw^2}{\cw^2}=
-\frac{1}{2} \besw_{B}\lu+
\de Z^{\elm}_e, \nl
\frac{\de g_2}{g_2}&=&\de Z_e +\frac{1}{2}\frac{\cw^2}{\sw^2}\frac{\delta \cw^2}{\cw^2}=
-\frac{1}{2} \besw_{W}\lu+
\de Z^{\elm}_e,
\eeqar
where we have used the relations \refeq{betarelations}.

\subsection*{Yukawa-coupling renormalization}

In the on-shell scheme, the renormalization of the top-quark mass is
given by
\beq
\de \Mt =
\frac{\Mt}{2}\left[\Si^{t,\rL}(\Mt^2)+\Si^{t,\rR}(\Mt^2)+2\Si^{t,\rS}(\Mt^2)\right].
\eeq
This leads to the following logarithmic contributions
\newcommand{\Qt}{Q_{\Pt}}
\beq
\frac{\de \Mt}{\Mt} =
\biggl[\frac{1}{4\sw^2}+\frac{1}{8\sw^2\cw^2}+\frac{3}{2\cw^2}\Qt -\frac{3}{\cw^2}\Qt^2
+\frac{3}{8\sw^2}\frac{\Mt^2}{\MW^2}\biggr]\lu.
\eeq
Using \refeq{massCT}, the counterterm for $\rt$ reads
\beqar
\frac{\de\rt}{\rt} &=& \frac{\de \Mt}{\Mt}- \frac{1}{2}\frac{\de
  \MW^2}{\MW^2}\nl
&=&\frac{1}{2}\bew_{W}\lu 
+\biggl[-\frac{3}{4\sw^2}-\frac{3}{8\sw^2\cw^2}+\frac{3}{2\cw^2}\Qt-\frac{3}{\cw^2}\Qt^2\biggr]\lu
\nl
&&{}+\frac{3+2\NCt}{8\sw^2}\frac{\Mt^2}{\MW^2}\lu.
\eeqar
The counterterm to the top-quark Yukawa coupling 
\beq
\gt = \frac{e}{\sqrt{2}\sw}\rt
\eeq
is given by
\beq
\frac{\de\gt}{\gt} = \frac{1}{2}\Delta\al(\MW^2) - \frac{1}{2}\bew_{W}\lu
+\frac{\de\rt}{\rt}.
\eeq

\subsection*{Scalar-self-coupling renormalization}

In the on-shell scheme, the renormalization of the Higgs mass is
given by
\beq
\de \MH^2 = \Si^H(\MH^2),
\eeq
or in logarithmic accuracy
\beqar
\frac{\de \MH^2}{\MH^2} &=&
\frac{1}{2\sw^2}\biggl[\frac{9\MW^2}{\MH^2}\left(1+\frac{1}{2\cw^4}\right)
-\frac{3}{2}\left(1+\frac{1}{2\cw^2}\right) + \frac{15}{4}\frac{\MH^2}{\MW^2}\biggr]\lu\nl
&&{}+\frac{\NCt}{2\sw^2}\frac{\Mt^2}{\MW^2}\left(1-6\frac{\Mt^2}{\MH^2}\right)
\lu.
\eeqar
The renormalization of the  scalar self couplings gets an extra
contribution from the tadpole renormalization
(\cf\citere{bfm})
\beqar
\de t &=& -T =
\frac{1}{e\sw\MW}\biggl[-\frac{3}{2}\MW^2\left(\frac{\MZ^2}{\cw^2}+2\MW^2\right)
\nl &&{}
-\frac{\MH^2}{4}(2\MW^2+\MZ^2+3\MH^2)+2\NCt\Mt^4\biggr]\lu.
\eeqar
Including this in the renormalization of $\rH$ and using
\refeq{massCT}, we find the effective counterterm
\beqar
\frac{\de\rH^{\eff}}{\rH} &=& \frac{\de \MH^2}{\MH^2}
 - \frac{\de \MW^2}{\MW^2}+\frac{e}{2\sw}\frac{\de t}{\MW\MH^2}\nl
&=& \bew_{W}\lu +
\frac{3}{2\sw^2}\biggl[\frac{\MW^2}{\MH^2}\left(2+\frac{1}{\cw^4}\right)
-\left(2+\frac{1}{\cw^2}\right) + \frac{\MH^2}{\MW^2}\biggr]\lu\nl
&&{}
+\frac{\NCt}{\sw^2}\frac{\Mt^2}{\MW^2}\left(1-2\frac{\Mt^2}{\MH^2}\right)\lu.
\eeqar
The counterterm to the scalar-self coupling
\beq
\gH = \frac{e^2}{{2}\sw^2}\rH
\eeq
is given by
\beq
\frac{\de\gH}{\gH} =
\Delta\al(\MW^2) - \bew_{W}\lu +\frac{\de\rH^{\eff}}{\rH}.
\eeq

The logarithms resulting from parameter renormalization are those that
determine the running of the couplings.

\section{Applications to simple processes}
\label{se:applicat}
In this section, the above results for Sudakov DL, collinear or soft SL, 
and PR corrections are applied to simple processes. We discuss relative
corrections to polarized Born amplitudes,
\beq \label{relco}
\de_{A\rightarrow B}=\frac{\de \M^{A\rightarrow B}}{\M_0^{A\rightarrow B}}.
\eeq  
Note that the corrections to the cross sections are twice as large.
The complete logarithmic corrections are presented in analytic form.
The numerical results are given for the coefficients of the genuine
electroweak (ew) logarithms. These are obtained by omitting the 
pure electromagnetic contributions that result from the gap between the
electromagnetic and the weak scale. Accordingly they include the
symmetric-electroweak contributions and the 
$\ls$  terms originating from $\PZ$-boson loops in \refeq{deSC}.
In order to keep track of the origin of the various  
$\ls$ terms, 
we introduce different
subscripts: collinear, Yukawa, PR contributions, 
and the $\PZ$-boson contributions from \refeq{deSC} are denoted by
$\lsl$, $\lYuk$, $\lpr$, and $\lZ$ respectively.
The numerical results have
been obtained using the following values for the physical parameters:
\begin{displaymath}
\MW=80.35 \GeV,\qquad \MZ=91.1867 \GeV,\qquad  \Mt=175 \GeV,
\end{displaymath}
\beq
\alpha=\frac{1}{137.036},\qquad \sw^2=1-\frac{\MW^2}{\MZ^2}\approx0.22356.
\eeq

\subsection{Four-fermion neutral-current processes}
The Sudakov DL corrections \refeq{deSC} and the collinear or soft SL
corrections \refeq{deccfer} depend only on the quantum numbers of the
external legs, and can be applied to 4-fermion processes in a
universal way. However, we are interested also in the $\SS$ and $\pre$
corrections, which depend on the specific properties of the process. A
general description of these corrections requires a decomposition of
the Born matrix element into neutral-current (NC) and charged-current
(CC) contributions.  In order to simplify the discussion we restrict
ourselves to pure NC transitions.  To simplify notation, we consider
processes involving a lepton--antilepton and a quark--antiquark pair.
However, our analysis applies to the more general case of two
fermion--antifermion pairs of different isospin doublets.
The 4 external states and their momenta 
are chosen to be incoming, so that the process reads 
\beq \label{4fprocess} 
\bar{l}^\kappa_\si l^\kappa_\si q^{\la}_{\rho} \bar{q}^{\la}_{\rho}\rightarrow 0, 
\eeq
where $\kappa,\la=\rR,\rL$ are the chiralities and $\si,\rho=\pm$  the isospin indices. All formulas for the $4\rightarrow 0$ process \refeq{4fprocess}
are expressed in terms of the particle eigenvalues $I^\NB_{l^\kappa_\si}$, $I^\NB_{q^{\la}_{\rho}}$. 

\newcommand{\NC}{R}
\newcommand{\NCew}{\NC_{\Pe^-_\kappa\PW^-_\la}}
\newcommand{\NCep}{\NC_{\Pe^-_\kappa\phi^-}}
\newcommand{\deNClq}{\Delta_{l^\kappa_\si q^\la_\rho}}
\newcommand{\deNCew}{\Delta_{\Pe^-_\kappa\PW^-_\rT}}
\newcommand{\deNCep}{\Delta_{\Pe^-_\kappa\phi^-}}
\newcommand{\NClq}{\NC_{l^\kappa_\si q^\la_\rho}}
\newcommand{\NCll}{\NC_{l^\kappa_\si l^\kappa_\si}}
\newcommand{\NCqq}{\NC_{q^\la_\rho q^\la_\rho}}
\newcommand{\NClmq}{\NC_{l^\kappa_{-\si} q^\la_\rho}}
\newcommand{\NClqm}{\NC_{l^\kappa_\si q^\la_{-\rho}}}

In the high-energy limit, the Born amplitude is given by
\beq \label{born4fnc} 
\M_{0}^{\bar{l}^\kappa_\si l^\kappa_\si q^{\la}_{\rho} \bar{q}^{\la}_{\rho}}
=e^2\NClq \frac{A_{12}}{r_{12}},
\eeq
where
\beq \label{NCcorr}
\NC_{\phi_i\phi_k}:=\sum_{\NB=A,Z}I^\NB_{\phi_i}I^\NB_{\phi_k}=\frac{1}{4\cw^2} Y_{\phi_i} Y_{\phi_k}+\frac{1}{\sw^2} T^3_{\phi_i}T^3_{\phi_k},
\eeq
and terms of order $\MZ^2/r_{12}$, originating from the difference between the photon and the Z-boson mass, are neglected. 
Note that \refeq{NCcorr} and the following formulas have an
important chirality
dependence, owing to the different values of the group-theoretical
operators in the representations for right-handed and left-handed
fermions.

The Sudakov soft--collinear corrections give according to \refeq{deSC}
the leading contribution
\beq
\de^{\SC}_{\bar{l}^\kappa_\si l^\kappa_\si q^{\la}_{\rho} \bar{q}^{\la}_{\rho}}
=-\sum_{f^\mu_\tau=l^\kappa_\si,q^{\la}_\rho} \left[\cew_{f^\mu}\Ls-2(I^Z_{f^\mu_\tau})^2 \log{\frac{\MZ^2}{\MW^2}}\,\lZ
+Q_{f_\tau}^2\Lemftau \right].
\eeq

The angular-dependent $\SS$ corrections are obtained from
\refeq{4fsubdl}.  The contribution of the neutral gauge bosons $\NB=A,Z$
is diagonal in the $\SUtwo$ indices, and factorizes into
the Born matrix element \refeq{born4fnc} times the relative correction
\beqar \label{4fncss} 
\sum_{\NB=A,Z}\de^{\NB,\SS}_{\bar{l}^\kappa_\si l^\kappa_\si q^{\la}_{\rho} \bar{q}^{\la}_{\rho}}
&=&-2\ls\left\{(\NCll+\NCqq)\log{\frac{|r_{12}|}{s}}
+2\NClq\log{\frac{|r_{13}|}{|r_{14}|}}
\right\}\nl
&&{}-2\lWla\left[(Q_{l_\si}^2+Q_{q_\rho}^2)\log{\frac{|r_{12}|}{s}}+2Q_{l_\si}Q_{q_\rho}\log{\frac{|r_{13}|}{|r_{14}|}}\right],
\eeqar
where $I^\NB_{\bar f \bar f} = -I^\NB_{ff}$ has been used and 
terms involving $\lWZ$ have been omitted. The contribution 
of the charged gauge bosons to \refeq{4fsubdl} gives  
\beqar \label{4fncss2} 
\sum_{\GB_a=W^\pm}\de^{\GB_a,\SS} \M^{\bar{l}^\kappa_\si l^\kappa_\si q^{\la}_{\rho} \bar{q}^{\la}_{\rho}}&=&-\frac{1}{\sw^2}\ls\left\{\left(
\de_{\kappa\rL}\M_0^{\bar{l}^\kappa_{-\si} l^\kappa_{-\si} q^{\la}_{\rho} \bar{q}^{\la}_{\rho}}
+\de_{\la\rL}\M_0^{\bar{l}^\kappa_\si l^\kappa_{\si} q^{\la}_{-\rho} \bar{q}^{\la}_{-\rho}}
\right)\log{\frac{|r_{12}|}{s}}\right.\nl
&&\left.{}+\de_{\kappa\rL}\de_{\la\rL}\left[\de_{\si\rho}\left(
\M_0^{\bar{l}^\kappa_{-\si} l^\kappa_\si q^{\la}_{-\rho} \bar{q}^{\la}_{\rho}}
+\M_0^{\bar{l}^\kappa_\si l^\kappa_{-\si} q^{\la}_{\rho} \bar{q}^{\la}_{-\rho}}
\right)\log{\frac{|r_{13}|}{s}}
\right.\right.\nl &&\left.\left.{}
-\de_{-\si\rho}\left(
\M_0^{\bar{l}^\kappa_{-\si} l^\kappa_\si q^{\la}_{\rho} \bar{q}^{\la}_{-\rho}}
+\M_0^{\bar{l}^\kappa_\si l^\kappa_{-\si} q^{\la}_{-\rho} \bar{q}^{\la}_{\rho}}
\right)\log{\frac{|r_{14}|}{s}}
\right]\right\},
\eeqar
where the non-diagonal couplings \refeq{ferpmcoup} have been used. On
the left-hand side, the $\SUtwo$-transformed Born matrix elements
involving the isospin partners $l^\kappa_{-\si}$, $q^\la_{-\rho}$,
have to be evaluated explicitly. The NC matrix elements (first line) are
obtained from \refeq{born4fnc}, and for the CC amplitudes
we find up to mass-suppressed terms using the non-diagonal couplings
\refeq{ferpmcoup},
\beq
\M_0^{\bar{l}^\kappa_{\si'} l^\kappa_{-\si'} q^{\la}_{\rho'} \bar{q}^{\la}_{-\rho'}}=\frac{e^2}{2\sw^2}\frac{A_{12}}{r_{12}}.
\eeq
Then, dividing \refeq{4fncss2} by the Born matrix element, we obtain
the relative correction
\beqar  \label{4fncss3} 
\sum_{\GB_a=W^\pm}\de^{\GB_a,\SS}_{\bar{l}^\kappa_\si l^\kappa_\si q^{\la}_{\rho} \bar{q}^{\la}_{\rho}} 
= -\frac{1}{\sw^2\NClq}\,\ls&&\left\{\left(\de_{\kappa\rL}\NClmq+ \de_{\la\rL}\NClqm\right)\log{\frac{|r_{12}|}{s}}\right.\nl
&&\left.{}+\frac{\de_{\kappa\rL}\de_{\la\rL}}{\sw^2}\left[\de_{\si\rho}\log{\frac{|r_{13}|}{s}}-\de_{-\si\rho}\log{\frac{|r_{14}|}{s}}\right]\right\}.
\eeqar

The angular-dependent corrections for $2\rightarrow 2$ processes, like
those depicted in \reffi{ffborn}, are directly obtained from
\refeq{4fncss} and \refeq{4fncss3} by substituting the invariants
$r_{kl}$ 
by 
the corresponding Mandelstam variables $s,t,u$.
For the $s$-channel processes
$\bar{l}^\kappa_{\si} l^\kappa_\si \rightarrow \bar{q}^{\la}_{\rho}
q^{\la}_{\rho}$, we have to substitute $r_{12}=s,r_{13}=t,r_{14}=u$,
and the $\SS$ corrections simplify to
\beqar \label{4fsSS} 
\de^\SS_{\bar{l}^\kappa_\si l^\kappa_\si \rightarrow \bar{q}^{\la}_{\rho} q^{\la}_{\rho}}&=&- \ls\left[4\NClq\ltu+\frac{\de_{\kappa\rL}\de_{\la\rL}}{\sw^4\NClq}\left(\de_{\si\rho}\lts-\de_{-\si\rho}\lus\right) \right]\nl
&&{}-4 Q_{l_\si}Q_{q_\rho}\lWla\ltu.
\eeqar
If one subtracts the photonic contributions from \refeq{4fsSS} one
finds agreement with eq.~(50) of \citere{Ku2}.  For the $t$-channel
processes $ \bar{q}^{\la}_{\rho} l^\kappa_\si \rightarrow
\bar{q}^{\la}_{\rho}l^\kappa_{\si}$, the substitution reads
$r_{12}=t,r_{13}=s,r_{14}=u$, whereas for $ q^{\la}_{\rho}
l^\kappa_\si \rightarrow q^{\la}_{\rho}l^\kappa_{\si}$ one has to
choose $r_{12}=t,r_{13}=u,r_{14}=s$.

\begin{figure}
\begin{center}
\begin{picture}(300,80)
\put(0,0){
\begin{picture}(120,80)
\ArrowLine(20,30)(0,60)
\ArrowLine(0,0)(20,30)
\Vertex(20,30){2}
\Photon(20,30)(80,30){2}{6}
\Vertex(80,30){2}
\ArrowLine(80,30)(100,0)
\ArrowLine(100,60)(80,30)
\Text(-5,0)[r]{$l^\kappa_\si$}
\Text(-5,60)[r]{$\bar l^\kappa_\si$}
\Text(105,0)[l]{$q^\la_\rho$}
\Text(105,60)[l]{$\bar q^\la_\rho$}
\Text(50,35)[b]{$A,Z$}
\end{picture}}
\put(180,0){
\begin{picture}(120,80)
\ArrowLine(40,45)(10,60)
\Photon(40,15)(40,45){2}{3}
\ArrowLine(10,0)(40,15)
\Vertex(40,15){2}
\Vertex(40,45){2}
\ArrowLine(40,15)(70,0)
\ArrowLine(70,60)(40,45)
\Text(5,0)[r]{$l^\kappa_\si$}
\Text(5,60)[r]{$\bar q^\la_\rho$}
\Text(45,30)[l]{$A,Z$}
\Text(75,0)[l]{$l^\kappa_\si$}
\Text(75,60)[l]{$\bar q^\la_\rho$}
\end{picture}}
\end{picture}
\end{center}
\caption{Lowest-order diagrams for
  $ \bar l^\kappa_\si l^\kappa_\si \to  \bar q^\la_\rho q^\la_\rho$ and $ \bar q^\la_\rho l^\kappa_\si \to  \bar q^\la_\rho l^\kappa_\si$}
\label{ffborn}
\end{figure}

The collinear or soft SL contributions \refeq{deccfer} give  
\beqar
\de^{\cc}_{\bar{l}^\kappa_\si l^\kappa_\si q^{\la}_{\rho} \bar{q}^{\la}_{\rho}}&=&\sum_{f^\mu_\tau=l^\kappa_\si,q^\la_\rho} \left[3 \cew_{f^\mu}\lsl -\frac{1}{4\sw^2}\left((1+\delta_{\mu R})\frac{m_{f_\tau}^2}{\MW^2}+\delta_{\mu L}\frac{m_{f_{-\tau}}^2}{\MW^2}\right)\lYuk \right.\nl
&& \hspace{1.2cm}\left. {}+2Q_{f_\tau}^2\lemftau \right],
\eeqar
and the Yukawa contribution depends on the chiralities $\mu$ and on the masses  of the fermions  $f^\mu_\tau$ and their isospin partners $f^\mu_{-\tau}$.

The PR logarithms for NC processes are obtained from the
renormalization of the electric charge and the weak mixing angle in the
Born amplitude \refeq{born4fnc}. Using \refeq{weinbergrenorm} and
\refeq{chargerenorm} this gives the relative correction
\beq
\de^\pre_{\bar{l}^\kappa_\si l^\kappa_\si q^{\la}_{\rho} \bar{q}^{\la}_{\rho}}=\left[\frac{\sw}{\cw}\bew_{AZ}\deNClq-\bew_{AA}\right]\lpr+2\de Z_e^\elm,
\eeq
where
\beq \label{NCren}
\Delta_{\phi_i\phi_k}:=\frac{-\frac{1}{4\cw^2}Y_{\phi_i}
  Y_{\phi_k}+\frac{\cw^2}{\sw^4}
  T^3_{\phi_i}T^3_{\phi_k}}{\NC_{\phi_i\phi_k}}
\eeq
gives a chirality-dependent contribution owing to mixing-angle
renormalization of \refeq{NCcorr}, and $\bew_{AA}$ represents the
universal contribution of electric charge renormalization.  

In order to give an impression of the size 
of the  genuine electroweak part of
the corrections, we consider the relative corrections $\de^{\kappa_\Pe
  \kappa_f,\sew}_{\Pep \Pem \rightarrow \bar{f}f}$ to NC processes
$\Pep\Pem\rightarrow \bar{f}f$ with chiralities
$\kappa_\Pe,\kappa_f=\rR$ or $\rL$, and give the numerical
coefficients of the electroweak  logarithms for the cases $f=\mu,\Pt,\Pb$. 
For muon-pair production we have
\newcommand{\eemumuRR}{\de^{\rR\rR,\sew}_{\Pep \Pem \rightarrow \mu^+\mu^-}}
\newcommand{\eemumuRL}{\de^{\rR\rL,\sew}_{\Pep \Pem \rightarrow \mu^+\mu^-}}
\newcommand{\eemumuLR}{\de^{\rL\rR,\sew}_{\Pep \Pem \rightarrow \mu^+\mu^-}}
\newcommand{\eemumuLL}{\de^{\rL\rL,\sew}_{\Pep \Pem \rightarrow \mu^+\mu^-}}
\beqar
\eemumuRR &=&   -2.58\,\Ls-5.15\left(\ltu\right)\ls+0.29\,\lZ+7.73\,\lsl+8.80\,\lpr   ,\nl
\eemumuRL &=&   -4.96\,\Ls-2.58\left(\ltu\right)\ls+0.37\,\lZ+14.9\,\lsl+8.80\,\lpr   ,\nl
\eemumuLL &=&   -7.35\,\Ls-\left(5.76\ltu+13.9\lts\right)\ls+0.45\,\lZ\nl
&&{}+22.1\,\lsl-9.03\,\lpr,
\eeqar
and $\eemumuLR =\eemumuRL$. For top-quark-pair production we find
\newcommand{\eettRR}{\de^{\rR\rR,\sew}_{\Pep \Pem \rightarrow  \bar{\Pt}\Pt}}
\newcommand{\eettRL}{\de^{\rR\rL,\sew}_{\Pep \Pem \rightarrow  \bar{\Pt}\Pt}}
\newcommand{\eettLR}{\de^{\rL\rR,\sew}_{\Pep \Pem \rightarrow  \bar{\Pt}\Pt}}
\newcommand{\eettLL}{\de^{\rL\rL,\sew}_{\Pep \Pem \rightarrow  \bar{\Pt}\Pt}}
\beqar \label{toppair}
\eettRR &=&   -1.86\,\Ls+3.43\left(\ltu\right)\ls+0.21\,\lZ+5.58\,\lsl-10.6\,\lYuk+8.80\,\lpr   ,\nl
\eettRL &=&   -4.68\,\Ls+0.86\left(\ltu\right)\ls+0.50\,\lZ+14.0\,\lsl-5.30\,\lYuk+8.80\,\lpr   ,\nl
\eettLR &=&   -4.25\,\Ls+1.72\left(\ltu\right)\ls+0.29\,\lZ+12.7\,\lsl-10.6\,\lYuk+8.80\,\lpr   ,\nl
\eettLL &=&   -7.07\,\Ls+\left(4.90\ltu-16.3\lus\right)\ls+0.58\,\lZ \nl
&&{}+21.2\,\lsl-5.30\,\lYuk-12.2\,\lpr,
\eeqar
and for bottom-quark-pair production we obtain
\newcommand{\eebbRR}{\de^{\rR\rR,\sew}_{\Pep \Pem \rightarrow \bar{\Pb}\Pb}}
\newcommand{\eebbRL}{\de^{\rR\rL,\sew}_{\Pep \Pem \rightarrow \bar{\Pb}\Pb}}
\newcommand{\eebbLR}{\de^{\rL\rR,\sew}_{\Pep \Pem \rightarrow \bar{\Pb}\Pb}}
\newcommand{\eebbLL}{\de^{\rL\rL,\sew}_{\Pep \Pem \rightarrow \bar{\Pb}\Pb}}
\beqar \label{bpair}
\eebbRR &=&   -1.43\,\Ls-1.72\left(\ltu\right)\ls+0.16\,\lZ+4.29\,\lsl+8.80\,\lpr   ,\nl
\eebbRL &=&   -4.68\,\Ls+0.86\left(\ltu\right)\ls+0.67\,\lZ+14.0\,\lsl-5.30\,\lYuk+8.80\,\lpr   ,\nl
\eebbLR &=&   -3.82\,\Ls-0.86\left(\ltu\right)\ls+0.24\,\lZ+11.5\,\lsl+8.80\,\lpr   ,\nl
\eebbLL &=&   -7.07\,\Ls-\left(4.04\ltu+19.8\lts\right)\ls+0.75\,\lZ \nl
&&{}+21.2\,\lsl-5.30\,\lYuk-16.6\,\lpr.
\eeqar 
The Mandelstam variables are defined
as usual, \ie
$s=(p_\Pep+p_\Pem)^2$, $t=(p_\Pep-p_{\bar{f}})^2$ and $u=(p_\Pep-p_{f})^2$.
Note that the corrections to light quark-pair production $f=\Pu,\Pc\,
(\Pd,\Ps)$ are obtained from the results for heavy quarks $f=\Pt\,
(\Pb)$ by omitting the Yukawa contributions. Independently of the
process and of the 
chirality,
the DL and SL terms appear in the
combination $(-\Ls+3\lsl)$, so that the negative DL contribution
becomes dominating only above $400 \GeV$, and at $\sqrt{s}=1 \TeV$ the
cancellation between SL and DL corrections is still important. The
$\SUtwo$ interaction, which is stronger than the $\Uone$ interaction, 
generates large corrections for left-handed fermions.  Also the PR
logarithms show a strong chirality dependence: the RR and RL
transitions receive positive corrections from the running of the
abelian $\Uone$ coupling, whereas the LL transition is dominated by
the non-abelian $\SUtwo$ interaction and receives negative PR
corrections.

\subsection{Production of $\PW$-boson pairs in $\Pep\Pem$ annihilation}\label{eeWW}
We consider the polarized scattering process\footnote{The momenta and
  fields of the initial states are incoming, and those of the
  final states are outgoing.}  
$\Pe^+_\kappa\Pe^-_\kappa \rightarrow \PW^+_{\la_+} \PW^-_{\la_-}$, where
$\kappa=\rR,\rL$ is the
electron chirality, and $\la_\pm=0,\pm$ represent the gauge-boson
helicities. In the high-energy limit only the following helicity
combinations are non-suppressed 
\cite{eeWWhe,Denn1}: the purely longitudinal final state
$(\la_+,\la_-)=(0,0)$, which we denote by $(\la_+,\la_-)=(\rL,\rL)$,
and the purely transverse and opposite final state  $(\la_+,\la_-)=(\pm,\mp)$,
which we denote by $(\la_+,\la_-)=(\rT,\rT)$. All these final states,
can be written as $(\la_+,\la_-)=(\la,-\la)$.
The Mandelstam
variables are $s=(p_\Pep+p_\Pem)^2$, $t=(p_\Pep-p_\PWp)^2\sim
-s(1-\cos{\theta})/2$, and $u=(p_\Pep-p_\PWm)^2\sim
-s(1+\cos{\theta})/2$, where $\theta$ is the angle between $\Pep$ and
$\PWp$.  The Born amplitude gets contributions of the $s$- and
$t$-channel diagrams in \reffi{WWborn} and reads
\beqar \label{borneeww} 
\M_{0}^{ \Pe^+_\kappa\Pe^-_\kappa \rightarrow \PW^+_\rL\PW^-_\rL}&=&
e^2\NCep \frac{A_s}{s},\nl 
\M_{0}^{\Pe^+_\rL\Pe^-_\rL \rightarrow
  \PW^+_\rT\PW^-_\rT}&=&\frac{e^2}{2\sw^2}\frac{A_t}{t} 
\eeqar
up to terms of order $\MW^2/s$, where $\NC$ is defined in \refeq{NCcorr}.
The amplitude involving
longitudinal gauge bosons $\PW_\rL$ is expressed by the amplitude
involving Goldstone bosons $\phi^\pm$ and is dominated by the
$s$-channel exchange of neutral gauge bosons. The
amplitude for transverse gauge-boson production is dominated by the
$t$-channel contribution, which involves only the $\SUtwo$ interaction.
Therefore, it is non-vanishing only for left-handed electrons in the
initial state.
\begin{figure}
\begin{center}
\begin{picture}(300,80)
\put(0,0){
\begin{picture}(120,80)
\ArrowLine(20,30)(0,60)
\ArrowLine(0,0)(20,30)
\Vertex(20,30){2}
\Photon(20,30)(80,30){2}{6}
\Vertex(80,30){2}
\DashLine(100,0)(80,30){2}
\DashLine(80,30)(100,60){2}
\Text(-5,0)[r]{$\Pem$}
\Text(-5,60)[r]{$\Pep$}
\Text(105,0)[l]{$\phi^-$}
\Text(105,60)[l]{$\phi^+$}
\Text(50,35)[b]{$A,Z$}
\end{picture}}
\put(180,0){
\begin{picture}(120,80)
\ArrowLine(40,45)(10,60)
\ArrowLine(40,15)(40,45)
\ArrowLine(10,0)(40,15)
\Vertex(40,15){2}
\Vertex(40,45){2}
\Photon(70,0)(40,15){2}{3}
\Photon(40,45)(70,60){-2}{3}
\Text(5,0)[r]{$\Pem$}
\Text(5,60)[r]{$\Pep$}
\Text(47,30)[l]{$\nu_\Pe$}
\Text(75,0)[l]{$\PW^-_\rT$}
\Text(75,60)[l]{$\PW^+_\rT$}
\end{picture}}
\end{picture}
\end{center}
\caption{Dominant  lowest-order diagrams for
  $\Pe^+\Pe^-\to\phi^+\phi^-$ and $\Pe^+\Pe^-\to \PW^+_\rT \PW^-_\rT$}
\label{WWborn}
\end{figure}

The DL corrections read
\beq\label{SCWW}
\de^{\SC}_{\Pe^+_\kappa\Pe^-_\kappa \rightarrow \PW^+_\la\PW^-_{-\la}}=-\sum_{\varphi=\Pe^-_\kappa,\PW^-_\la} \left[\cew_\varphi\Ls-2(I^Z_\varphi)^2 \log{\frac{\MZ^2}{\MW^2}}\,\lZ
+Q_\varphi^2\Lemphi \right].
\eeq
Here and in the following formulas,
for longitudinally polarized gauge bosons $\PW^\pm_\rL$ the quantum
numbers of the Goldstone bosons $\phi^\pm$ have to be used.

The $\SS$ corrections are obtained by applying \refeq{4fsubdl} to the
crossing symmetric process $\Pe^+_\kappa\Pe^-_\kappa
\PW^-_\la\PW^+_{-\la}\rightarrow 0$.  The contribution of the neutral
gauge bosons $\NB=A,Z$ gives
\beqar 
\sum_{\NB=A,Z}\de^{\NB,\SS}_{\Pe^+_\kappa\Pe^-_\kappa \rightarrow \PW^+_\la\PW^-_{-\la}}
=-4\NCew\ls\ltu -4Q_{\Pem}Q_{\PW^-}\lWla\ltu,
\eeqar
and corresponds to the result \refeq{4fncss} for 4-fermion s-channel NC processes. 
The contribution of soft $\PW^\pm$ 
bosons to \refeq{4fsubdl} yields
\beqar \label{eewwpmssc}
\sum_{\GB_a=W^\pm}\de^{\GB_a,\SS} \M^{\Pe^+_\kappa\Pe^-_\kappa\phi^-\phi^+}
&=&-\frac{2\ls\de_{\kappa\rL}}{\sqrt{2}\sw}\sum_{S=H,\chi}\left[
I^+_S \M_0^{\bar{\nu}_\kappa\Pe^-_\kappa S \phi^+}
-I^-_S\M_0^{\Pe^+_\kappa\nu_\kappa\phi^-S}
\right]
\log{\frac{|t|}{s}},\nl
\sum_{\GB_a=W^\pm}\de^{\GB_a,\SS} \M^{\Pe^+_\rL\Pe^-_\rL\PW_\rT^-\PW_\rT^+}
&=&-\frac{2\ls}{\sqrt{2}\sw}\sum_{\NB=A,Z}\left[
I^+_\NB\M_0^{\bar{\nu}_\rL\Pe^-_\rL \NB_\rT \PW_\rT^+}
-I^-_\NB\M_0^{\Pe^+_\rL\nu_\rL\PW_\rT^-\NB_\rT}
\right]
\log{\frac{|t|}{s}},\nl
\eeqar
where, depending on the polarization of the final states, one has to
use the non-diagonal $\PW^\pm$ couplings to Goldstone bosons
($I^\pm_S$) defined in \refeq{scapmcoup} or the $\PW^\pm$ couplings
to gauge bosons ($I^\pm_\NB$) defined in \refeq{gaupmcoup}.  The
$\SUtwo$-transformed Born matrix elements on the left-hand side of
\refeq{eewwpmssc} have to be evaluated explicitly. For Goldstone
bosons, we have $s$-channel CC amplitudes
\beq
\M_0^{\bar{\nu}_\kappa\Pe^-_\kappa S \phi^+}=-e^2I^-_S\frac{\de_{\kappa\rL}}{\sqrt{2}\sw} \frac{A_s}{s},\qquad
\M_0^{\Pe^+_\kappa\nu_\kappa\phi^-S}=e^2I^+_S\frac{\de_{\kappa\rL} }{\sqrt{2}\sw} \frac{A_s}{s},
\eeq
similar to the NC Born amplitude in \refeq{borneeww}, whereas for
transverse gauge bosons we have
\beq \label{borntransf}
\M_0^{\bar{\nu}_\kappa\Pe^-_\kappa \NB_\rT \PW_\rT^+}=
\M_0^{\Pe^+_\kappa\nu_\kappa\PW_\rT^-\NB_\rT}=
e^2\frac{\de_{\kappa\rL}}{\sqrt{2}\sw} \left(I^\NB_{\nu_\kappa}\frac{A_t}{t}+I^\NB_{e^-_\kappa}\frac{A_u}{u}\right),
\eeq
where  $A_t=A_u$ 
up to mass-suppressed contributions. In contrast to
\refeq{borneeww}, the transformed amplitude \refeq{borntransf}
receives contributions from both $t$ and $u$ channels.  Expressing
\refeq{eewwpmssc} as relative corrections to the Born matrix elements
we obtain
\beqar
\sum_{\GB_a=W^\pm}\de^{\GB_a,\SS}_{\Pe^+_\kappa\Pe^-_\kappa\rightarrow \PW_\rL^+\PW_\rL^-}
&=&-\ls\frac{\de_{\kappa\rL}}{\sw^4 \NC_{\Pe^-_\rL\phi^-}}\log{\frac{|t|}{s}},\nl
\sum_{\GB_a=W^\pm}\de^{\GB_a,\SS}_{\Pe^+_\rL\Pe^-_\rL\rightarrow \PW_\rT^+\PW_\rT^-}
&=&-\frac{2}{\sw^2}\left(1-\frac{t}{u}\right)\ls\log{\frac{|t|}{s}}.
\eeqar 

The SL corrections can be read off from \refeq{deccfer},
\refeq{deccWT}, and \refeq{longeq:coll},
\beqar
\de^\cc_{\Pe^+_\kappa\Pe^-_\kappa \rightarrow \PW^+_\rL\PW^-_\rL} &=& \left[3 \cew_{\Pe_\kappa}+ 4\cew_\Phi\right]\lsl-\frac{3}{2\sw^2}\frac{m^2_t}{\MW^2}\lYuk +\sum_{\varphi=\Pe,\PW}2\lemphi,\nl
\de^\cc_{\Pe^+_\rL\Pe^-_\rL \rightarrow \PW^+_\rT\PW^-_\rT} &=& \left[3 \cew_{\Pe_\rL}+\bew_{W}\right]\lsl+\sum_{\varphi=\Pe,\PW}2\lemphi.
\eeqar
Despite of their different origin, the $\lsl$ contributions for
longitudinal and transverse gauge bosons have similar numerical values
$4\cew_\Phi=14.707$ and $\bew_{W}=14.165$. 
The strong W-polarization dependence of $\de^\cc$ is due
to the large Yukawa contributions occurring only for longitudinal
gauge bosons.

The PR logarithms are obtained from the renormalization of
\refeq{borneeww} and read according to
\refeq{weinbergrenorm}--\refeq{gCTs}
\beqar\label{RGWW}
\de^\pre_{\Pe^+_\kappa\Pe^-_\kappa \rightarrow \PW^+_\rL\PW^-_\rL} &=&\left[\frac{\sw}{\cw}\bew_{AZ}\deNCep -\bew_{AA}\right]\lpr+2\de Z_e^\elm,\nl
\de^\pre_{\Pe^+_\rL\Pe^-_\rL \rightarrow \PW^+_\rT\PW^-_\rT} &=&-\bew_{W}\lpr+2\de Z_e^\elm,
\eeqar
where $\Delta$ is defined in \refeq{NCren}. 
Note that for transverse polarizations, the 
symmetric-electroweak 
parts
of the PR corrections 
($-\bew_W\lpr$) 
and the collinear SL corrections originating from external gauge bosons 
($\bew_W\lsl$) 
cancel. 
As illustrated in \refapp{app:transvRG}, this kind of cancellation
takes place for all processes with production of arbitrary many charged or
neutral transverse gauge bosons in fermion--antifermion annihilation.  

\begin{figure}
\centerline{
\setlength{\unitlength}{1cm}
\begin{picture}(10,8.3)
\put(0,0){\includegraphics{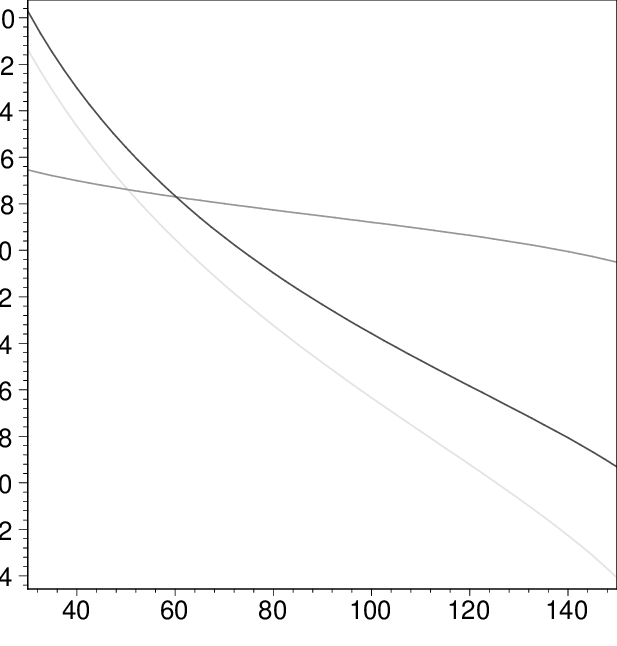}}
\put(2.5,0){\makebox(6,0.5)[b]{$\theta\,[^\circ]$}}
\put(-2.5,4){\makebox(1.5,1)[r]{$\delta^\sew\,[\%]$}}
\put(10,4.5){\makebox(1.5,1)[r]{$\rR\rL$}}
\put(10,2.0){\makebox(1.5,1)[r]{$\rL\rL$}}
\put(10,0.5){\makebox(1.5,1)[r]{$\rL\rT$}}
\end{picture}}
\caption[WWang]{Dependence of the electroweak correction factor
  $\de^{\sew}_{\Pe_\kappa^+\Pe_\kappa^-\rightarrow \PW_\la^+\PW_{-\la}^-}$ on
  the scattering angle $\theta$ at $\sqrt{s}=1\TeV$ for polarizations
  $\rR\rL$, $\rL\rL$, and $\rL\rT$} 
\label{plotWWan}
\end{figure}

\begin{figure}
  \centerline{ \setlength{\unitlength}{1cm}
\begin{picture}(10,8.3)
\put(0,0){\includegraphics{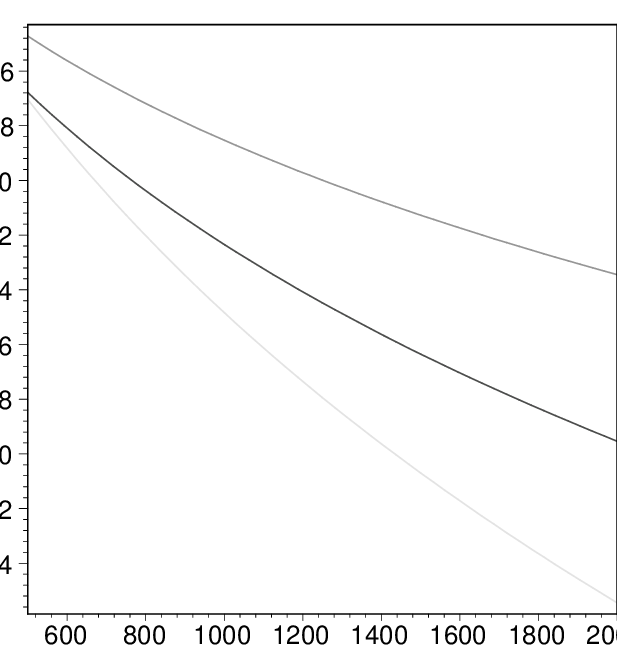}}
\put(2.5,0){\makebox(6,0.5)[b]{$\sqrt{s}\,[\GeV]$}}
\put(-2.5,4){\makebox(1.5,1)[r]{$\delta^\sew\,[\%]$}}
\put(10,4.5){\makebox(1.5,1)[r]{$\rR\rL$}}
\put(10,2.6){\makebox(1.5,1)[r]{$\rL\rL$}}
\put(10,0.8){\makebox(1.5,1)[r]{$\rL\rT$}}
\end{picture}}
\caption[WWang]{Dependence of the electroweak correction factor
  $\de^{\sew}_{\Pe_\kappa^+\Pe_\kappa^-\rightarrow \PW_\la^+\PW_{-\la}^-}$ on
  the energy  $\sqrt{s}$ at $\theta=90^\circ$ for polarizations
  $\rR\rL$, $\rL\rL$, and $\rL\rT$} 
\label{plotWWen}
\end{figure}

The results \refeq{SCWW}--\refeq{RGWW} can be compared with those of
\citere{eeWWhe}. After subtracting the real soft-photonic corrections
from the results of \citere{eeWWhe} we find complete agreement for the
logarithmic corrections.
The coefficients for the various 
electroweak logarithmic contributions to the relative
corrections 
$\de^{\kappa\la}_{\Pe^+\Pe^-\rightarrow \PW^+\PW^-}$ 
read
\newcommand{\eeWWLL}{\de^{\rL\rL,\sew}_{\Pe^+\Pe^-\rightarrow \PW^+\PW^-}}
\newcommand{\eeWWRL}{\de^{\rR\rL,\sew}_{\Pe^+\Pe^-\rightarrow \PW^+\PW^-}}
\newcommand{\eeWWLT}{\de^{\rL\rT,\sew}_{\Pe^+\Pe^-\rightarrow \PW^+\PW^-}}
\beqar
\eeWWLL &=&   -7.35\,\Ls-\left(5.76\ltu+13.9\lts\right)\ls+0.45\,\lZ \nl 
&&{}+25.7\,\lsl-31.8\,\lYuk-9.03\,\lpr   ,\nl
\eeWWRL &=&   -4.96\,\Ls-2.58\left(\ltu\right)\ls+0.37\,\lZ \nl &&
{}+18.6\,\lsl-31.8\,\lYuk+8.80\,\lpr   ,\nl
\eeWWLT &=&   -12.6\,\Ls-8.95\left[\ltu+\left(1-\frac{t}{u}\right)\lts\right]\ls+1.98\,\lZ \nl &&{}+25.2\,\lsl-14.2\,\lpr.
\eeqar 
Recall that the 
pure electromagnetic contributions have been omitted.
These correction factors are shown in  \reffis{plotWWan} and
\ref{plotWWen} as a function of the scattering angle and the energy, respectively.
If the electrons are left-handed, large negative DL and PR
corrections originate from the $\SUtwo$ interaction. Instead, for
right-handed electrons the DL corrections are smaller, and the PR
contribution is positive. 
For transverse \PW~bosons, there are no Yukawa contributions and the
other contributions are in general larger than for longitudinal
\PW~bosons.  Nevertheless, for energies around $1\TeV$, the
corrections are similar.
Finally, note that the angular-dependent contributions are
very important for the LL and LT corrections: at $\sqrt{s}\approx 1\TeV$
they vary from $+15\%$ to $-5\%$ for scattering angles
$30^\circ<\theta<150^\circ$, whereas the angular-dependent part of the
RL corrections remains between $\pm 2\%$.


\subsection{Production of neutral gauge-boson pairs in $\Pep\Pem$ annihilation}
We consider the polarized scattering process $\Pe^+_\kappa\Pe^-_\kappa
\rightarrow \NB_\rT^1 \NB_\rT^2$ with incoming electrons of chirality $\kappa=\rR,\rL$
and outgoing gauge bosons $\NB^k=\PA,\PZ$. The amplitude is
non-suppressed only for transverse and opposite gauge-boson
polarizations $(\la_1,\la_2)=(\pm,\mp)$ \cite{Denn2}. In lowest order
the $t$- and $u$-channel diagrams (\reffi{ZZborn}) yield
\beq \label{borneevv} 
\M_{0}^{ \Pe^+_\kappa\Pe^-_\kappa \rightarrow \NB_\rT^1\NB_\rT^2}= e^2 I^{\NB^1}_{\Pe^-_\kappa} I^{\NB^2}_{\Pe^-_\kappa}\left[ \frac{A_t}{t}+\frac{A_u}{u}\right]
\eeq
up to terms of order $\MW^2/s$, where the Mandelstam variables are
defined as in \refse{eeWW}. 
In the ultra-relativistic limit  the
amplitude is symmetric with respect to exchange of the gauge bosons,
and up to mass-suppressed contributions we have  
\beq \label{eevvchannel}
A_t=A_u.
\eeq 
\begin{figure}
\begin{center}
\begin{picture}(300,80)
\put(0,0){
\begin{picture}(120,80)
\ArrowLine(40,45)(10,60)
\ArrowLine(40,15)(40,45)
\ArrowLine(10,0)(40,15)
\Vertex(40,15){2}
\Vertex(40,45){2}
\Photon(70,0)(40,15){2}{3}
\Photon(40,45)(70,60){-2}{3}
\Text(5,0)[r]{$\Pem$}
\Text(5,60)[r]{$\Pep$}
\Text(75,0)[l]{$\NB^2$}
\Text(75,60)[l]{$\NB^1$}
\end{picture}}
\put(180,0){
\begin{picture}(120,80)
\ArrowLine(40,45)(10,60)
\ArrowLine(40,15)(40,45)
\ArrowLine(10,0)(40,15)
\Vertex(40,15){2}
\Vertex(40,45){2}
\Photon(70,0)(40,45){2}{5}
\Photon(40,15)(70,60){-2}{5}
\Text(5,0)[r]{$\Pem$}
\Text(5,60)[r]{$\Pep$}
\Text(75,0)[l]{$\NB^2$}
\Text(75,60)[l]{$\NB^1$}
\end{picture}}
\end{picture}
\end{center}
\caption{Lowest-order diagrams for $\Pe^+\Pe^-\to \NB^1\NB^2$}
\label{ZZborn}
\end{figure}

The DL corrections read [cf. \refeq{deSC}]
\beqar
\lefteqn{\de^{\SC}\M^{\Pe^+_\kappa\Pe^-_\kappa \rightarrow
    \NB_\rT^1\NB_\rT^2}=}\quad\nl
&&{}-\left[\cew_{\Pe_\kappa}\Ls-2(I^Z_{\Pe_\kappa})^2
    \log{\frac{\MZ^2}{\MW^2}}\,\lZ+\Leme
  \right]\M_{0}^{\Pe^+_\kappa\Pe^-_\kappa \rightarrow \NB_\rT^1\NB_\rT^2}\nl
  &&{}-\frac{1}{2}\left[\cew_{\NB^{'}\NB^1}\M_{0}^{\Pe^+_\kappa\Pe^-_\kappa
      \rightarrow
      \NB_\rT^{'}\NB_\rT^2}+\cew_{\NB^{'}\NB^2}\M_{0}^{\Pe^+_\kappa\Pe^-_\kappa
      \rightarrow \NB_\rT^1\NB_\rT^{'}} \right] \Ls
\eeqar
with a  non-diagonal contribution associated with the external neutral gauge bosons. Using
\beq \label{cewIid}
\cew_{\NB'\NB}I^{\NB'}=\frac{2}{\sw^2} U_{\NB W^3}(\thw) \tilde{I}^{W^3}=\frac{2}{\sw^2} U_{\NB W^3}(\thw)\frac{T^3}{\sw},
\eeq
where $U_{\NB \sNB}(\thw)$ is the Weinberg rotation defined in \refeq{weinbergrotation}, we can derive a correction relative to the Born matrix element,
\beqar \label{vvssc}
\de^{\SC}_{\Pe^+_\kappa\Pe^-_\kappa \rightarrow \NB_\rT^1\NB_\rT^2}&=&-\left[\cew_{\Pe_\kappa}\Ls-2(I^Z_{\Pe_\kappa})^2 \log{\frac{\MZ^2}{\MW^2}}\,\lZ+\Leme \right]\nl
&&{}-\frac{T^3_{\Pe^-_\kappa}}{\sw^3}\sum_{k=1,2} \frac{U_{\NB^k W^3}(\thw)}{I^{\NB^k}_{\Pe^-_\kappa}} \Ls.
\eeqar
Note that only the $\SUtwo$ component of the neutral gauge bosons is
self-interacting and can exchange soft gauge bosons. For this reason, only  left-handed electrons ($T^3\neq 0$) yield a contribution to \refeq{cewIid} and to the corresponding term in \refeq{vvssc}.

Angular-dependent logarithmic corrections  \refeq{4fsubdl}
arise only from the exchange of soft $\PW^\pm$ bosons between initial and final states, and with the non-diagonal couplings  \refeq{gaupmcoup}
\beqar \label{eevvpmssc}
\de^{\SS} \M^{\Pe^+_\kappa\Pe^-_\kappa \NB^1_\rT \NB^2_\rT}
=\frac{2\ls\de_{\kappa\rL}}{\sqrt{2}\sw}&&\left\{\left[
I^+_{\NB^1}\M_0^{\bar{\nu}_\kappa\Pe^-_\kappa \PW_\rT^+\NB^2_\rT}
-I^-_{\NB^2}\M_0^{\Pe^+_\kappa\nu_\kappa \NB^1_\rT\PW_\rT^-}
\right]\log{\frac{|t|}{s}}\right.\nl
&&\left.+\left[
I^+_{\NB^2}\M_0^{\bar{\nu}_\kappa\Pe^-_\kappa \NB^1_\rT\PW_\rT^+}
-I^-_{\NB^1}\M_0^{\Pe^+_\kappa\nu_\kappa \PW_\rT^-\NB^2_\rT}
\right]
\log{\frac{|u|}{s}}\right\}.
\eeqar
The $\SUtwo$-transformed Born matrix elements on the left-hand side
are given by
\beq
\M_0^{\bar{\nu}_\kappa\Pe^-_\kappa \PW_\rT^+ \NB_\rT}=
\M_0^{\Pe^+_\kappa\nu_\kappa \NB_\rT \PW_\rT^-}=
e^2\frac{\de_{\kappa\rL}}{\sqrt{2}\sw}
\left(I^\NB_{\Pe^-_\kappa}\frac{A_t}{t}+I^\NB_{\nu_\kappa}\frac{A_u}{u}\right)
\eeq
and by \refeq{borntransf} with $A_t=A_u$. Expressing the correction
\refeq{eevvpmssc} relative to the Born matrix element \refeq{borneevv},
we obtain
\beq 
\de^{\SS}_{\Pe^+_\kappa\Pe^-_\kappa\rightarrow \NB^1_\rT \NB^2_\rT}=
\frac{\de_{\kappa\rL}}{\sw^2}\ls \sum_{k=1}^2 \sum_{r=t,u} \frac{I^{\NB^k}_{\PWm}}{I^{\NB^k}_{\Pe^-_\rL}}\left(\frac{r'}{s}+\frac{r}{s}\frac{I^{\NB^{k'}}_{\nu_\rL}}{I^{\NB^{k'}}_{\Pe^-_\rL}}\right)\log{\frac{|r|}{s}},
\eeq
where $r'=(t,u)$ for  $r=(u,t)$, and  $k'=(1,2)$ for  $k=(2,1)$. 

\begin{figure}
\centerline{
\setlength{\unitlength}{1cm}
\begin{picture}(10,8.3)
\put(0,0){\includegraphics{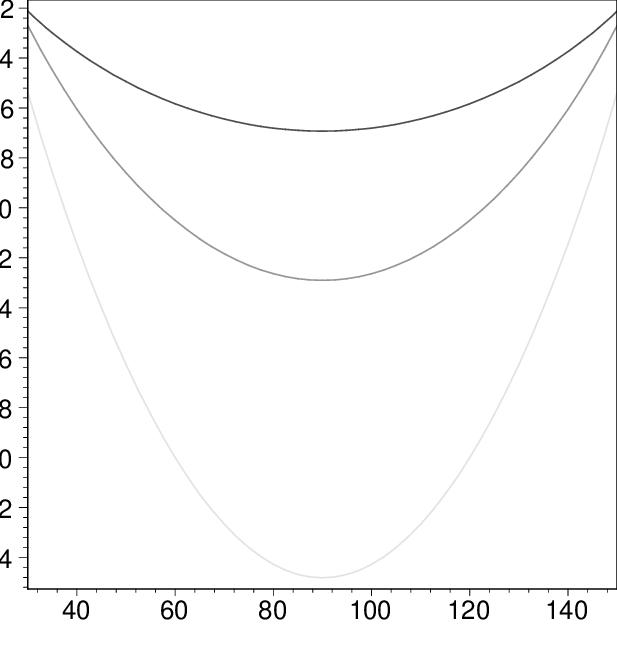}}
\put(2.5,0){\makebox(6,0.5)[b]{$\theta\,[^\circ]$}}
\put(-2.5,4){\makebox(1.5,1)[r]{$\delta^\sew\,[\%]$}}
\put(6,6.5){\makebox(1.5,1)[r]{$\PA\PA$}}
\put(6,5){\makebox(1.5,1)[r]{$\PA\PZ$}}
\put(6,2){\makebox(1.5,1)[r]{$\PZ\PZ$}}
\end{picture}}
\caption[AZLang]{Angular dependence of the electroweak corrections to
  $\Pe_\rL^+\Pe_\rL^-\rightarrow \PA\PA,\PA\PZ,\PZ\PZ$ at $\sqrt{s}=1\TeV$}
\label{plotAZan}
\end{figure}

\begin{figure}
  \centerline{ \setlength{\unitlength}{1cm}
\begin{picture}(10,8.3)
\put(0,0){\includegraphics{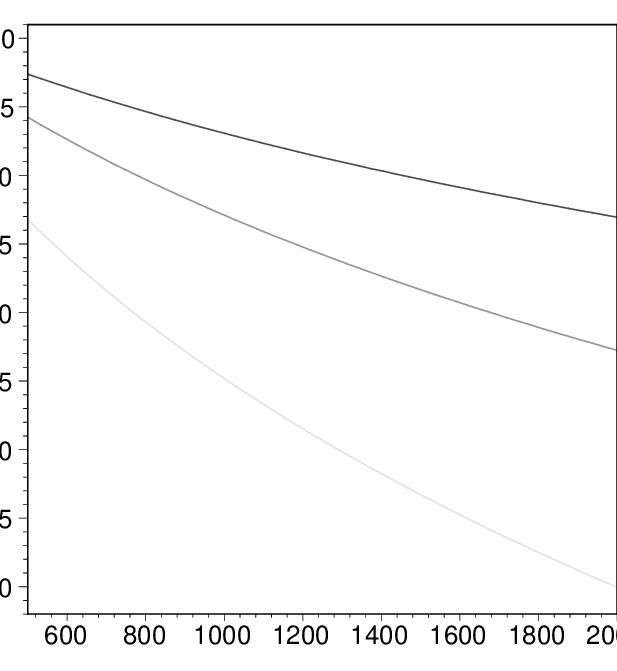}}
\put(2.5,0){\makebox(6,0.5)[b]{$\sqrt{s}\,[\GeV]$}}
\put(-2.5,4){\makebox(1.5,1)[r]{$\delta^\sew\,[\%]$}}
\put(6,6.5){\makebox(1.5,1)[r]{$\PA\PA$}}
\put(6,5){\makebox(1.5,1)[r]{$\PA\PZ$}}
\put(6,2.5){\makebox(1.5,1)[r]{$\PZ\PZ$}}
\end{picture}}
\caption[AZLen]{Energy dependence of the electroweak corrections to
  $\Pe_\rL^+\Pe_\rL^-\rightarrow \PA\PA,\PA\PZ,\PZ\PZ$ at $\theta=90^\circ$}
\label{plotAZen}
\end{figure}

Using \refeq{deccfer} and \refeq{deccVVT} we obtain for the SL
corrections relative to the Born matrix element 
\beq
\de^\cc_{\Pe^+_\kappa\Pe^-_\kappa \rightarrow \NB_\rT^1\NB_\rT^2}= 3\cew_{\Pe_\kappa}\lsl+2\leme+\de^\cc_{\NB_\rT^1}+\de^\cc_{\NB_\rT^2}
\eeq
with
\beqar
\de^\cc_A &:=&\de^\cc_{AA}(\GB_\rT)=\frac{1}{2}\bew_{AA}\lsl -\de Z^\elm_e,\nl
\de^\cc_Z &:=&\de^\cc_{ZZ}(\GB_\rT)+\de^\cc_{AZ}(\GB_\rT)\frac{\M_0^{\Pe^+_\kappa\Pe^-_\kappa \rightarrow A_\rT \NB_\rT}}{\M_0^{\Pe^+_\kappa\Pe^-_\kappa \rightarrow Z_\rT \NB_\rT}} 
=\left[\frac{1}{2}\bew_{ZZ}+\bew_{AZ}\frac{I^A_{\Pe^-_\kappa}}{I^Z_{\Pe^-_\kappa}}\right]\lsl.
\eeqar
The PR logarithms result from the renormalization of \refeq{borneevv}.
As shown in \refapp{app:transvRG}, they are opposite to the 
collinear SL corrections up to pure electromagnetic logarithms. 
Relative to the Born matrix element they read
\beq
\de^\pre_{\Pe^+_\kappa\Pe^-_\kappa \rightarrow \NB_\rT^1\NB_\rT^2}= \de^\pre_{\NB_\rT^1}+\de^\pre_{\NB_\rT^2}
\eeq
with 
\beq
\de^\pre_A :=-\de^\cc_A,\qquad
\de^\pre_Z :=-\de^\cc_Z +\de Z^\elm_e.
\eeq
For right-handed electrons, $\kappa=\rR$, the various 
electroweak
logarithmic contributions to the relative corrections
$\de^{\kappa\rT}_{\Pe^+\Pe^-\rightarrow \NB^1\NB^2}$ give
\newcommand{\eeAAL}{\de^{\rL\rT,\sew}_{\Pe^+\Pe^-\rightarrow \PA\PA}}
\newcommand{\eeAAR}{\de^{\rR\rT,\sew}_{\Pe^+\Pe^-\rightarrow \PA\PA}}
\newcommand{\eeAZL}{\de^{\rL\rT,\sew}_{\Pe^+\Pe^-\rightarrow \PA\PZ}}
\newcommand{\eeAZR}{\de^{\rR\rT,\sew}_{\Pe^+\Pe^-\rightarrow \PA\PZ}}
\newcommand{\eeZAL}{\de^{\rL\rT,\sew}_{\Pe^+\Pe^-\rightarrow \PZ\PA}}
\newcommand{\eeZAR}{\de^{\rR\rT,\sew}_{\Pe^+\Pe^-\rightarrow \PZ\PA}}
\newcommand{\eeZZL}{\de^{\rL\rT,\sew}_{\Pe^+\Pe^-\rightarrow \PZ\PZ}}
\newcommand{\eeZZR}{\de^{\rR\rT,\sew}_{\Pe^+\Pe^-\rightarrow \PZ\PZ}}
\newcommand{\UTTU}{F_1(t)}
\newcommand{\TTUU}{F_2(t)}
\beqar
\eeAAR &=&   -1.29\,\Ls+0.15\,\lZ+0.20\,\lsl+3.67\,\lpr   ,\nl
\eeAZR &=&   -1.29\,\Ls+0.15\,\lZ-11.3\,\lsl+15.1\,\lpr   ,\nl
\eeZZR &=&   -1.29\,\Ls+0.15\,\lZ-22.8\,\lsl+26.6\,\lpr    .
\eeqar
Note that there is no angular dependence. The PR contributions are
numerically compensated by the SL and DL Sudakov contributions, and at
$\sqrt{s}=1 \TeV$ the electroweak logarithmic corrections are less than
1$\%$. For left-handed electrons, we find
\beqar
\eeAAL &=&-8.15\,\Ls+8.95\UTTU\ls+0.22\,\lZ+7.36\,\lsl+3.67\,\lpr   ,\\
\eeAZL &=& -12.2\,\Ls+(17.0\UTTU-8.09\TTUU)\ls+0.22\,\lZ+28.1\,\lsl-17.1\,\lpr ,\nl
\eeZZL &=&-16.2\,\Ls+(25.1\UTTU-45.4\TTUU)\ls+0.22\,\lZ+48.9\,\lsl-37.9\,\lpr
\nn
\eeqar
with the $(t,u)$-symmetric angular-dependent functions
\beq
\UTTU:=\frac{u}{s}\lts+\frac{t}{s}\lus ,\qquad
\TTUU:= \frac{t}{s}\lts+\frac{u}{s}\lus.
\eeq
For left-handed electrons all contributions are larger than for
right-handed electrons owing to the $\SUtwo$ interaction. The
non-abelian effects are particularly strong for Z-boson-pair production
(see Figs. \ref{plotAZan}, \ref{plotAZen}), where the total
corrections are almost $-25\%$ for $\sqrt{s}=1 \TeV$ and
$\theta=90^\circ$. The angular-dependent contribution is
forward--backward symmetric, and for $\PZ\PZ$ production it
varies from $+15\%$ to $-5\%$ for scattering angles
$30^\circ<\theta<90^\circ$.


\section{Conclusion}

We have considered general electroweak processes at high energies.  We
have given recipes and explicit formulas for the extraction of the
one-loop leading electroweak logarithms. Like the well-known
soft--collinear double logarithms, also the collinear single
logarithms can be expressed as simple correction factors that are
associated with the external particles of the considered process. Up
to electromagnetic terms, the collinear SL corrections for external
longitudinal gauge bosons and for Higgs bosons are equal.  The
subleading single logarithms arising from the soft--collinear limit
are angular-dependent and can be associated to pairs of external
particles. Their evaluation requires in general all matrix elements
that are linked to the lowest-order matrix element via global $\SU(2)$
rotations.
Finally, the logarithms originating from coupling-constant
renormalization are associated with the explicit dependence of the
lowest-order matrix element on the coupling parameters.

Our results are applicable to general amplitudes that are not
mass-suppressed, as long as all invariants are large compared to the
masses. 
As illustration, we have applied our general results 
to fermion--antifermion production and the pair production of charged
and neutral gauge bosons.
For processes involving resonances, like in
$\Pep\Pem\to\PWp\PWm\to 4\Pf$, the corresponding invariants are
evidently not large and our results must be applied to the
subprocesses $\Pep\Pem\to\PWp\PWm$ and $\PWpm\to2\Pf$.

\section*{Acknowledgements}
We thank W. Beenakker, M. Melles, and A. Werthenbach for discussions. 


\appendix 
\section{Production of transverse gauge bosons in fermion--antifermion annihilation}
\label{app:transvRG}
For transverse W-pair production, we have observed that the symmetric
electroweak parts of the PR contributions and of the collinear SL
corrections originating from the external gauge bosons cancel exactly.
Here we illustrate how this cancellation takes place when arbitrarily
many charged or neutral gauge bosons are produced in
fermion--antifermion annihilation. To be specific, we consider
electron--positron annihilation into $n$ transverse charged
gauge-boson pairs and $m$ transverse neutral gauge bosons $\NB=\PA,\PZ$,
\beq
\Pe^+_\kappa \Pe^-_\kappa \rightarrow   \PW^+_{1,\rT}
\ldots \PW^+_{n,\rT}\PW^-_{1,\rT} \ldots \PW^-_{n,\rT} \NB_{1,\rT}\ldots  \NB_{m,\rT}.
\eeq
Collinear SL contributions give the correction factor 
\beqar
\de^{\cc} \M^{\Pep\Pem  \PW^+_{1} \ldots \PW^-_{n}\NB_1 \ldots \NB_m} &=& \left[2\de^\cc_{ee}(L_\kappa)+2n\de^\cc_{WW}(\GB_\rT)\right] \M^{\Pep\Pem \PW^+_{1} \ldots \PW^-_{n} \NB_1 \ldots \NB_m}\nl
&&{}+\sum_{k=1}^m \delta^\cc_{\NB'_k \NB_k}(\GB_\rT) \M_0^{\Pep\Pem \PW^+_{1} \ldots \PW^-_{n} \NB_1 \ldots \NB'_k \ldots \NB_{m}},
\eeqar
and owing to mixing 
  we have a
non-diagonal factor \refeq{deccVVT} in the neutral sector.

For the considered processes, it turns out that in the high-energy
limit the contribution of 
coupling-constant renormalization can be written as a sum over the
external gauge bosons. This can be easily shown, starting from the
unbroken phase. If one considers the production of $n$ W-boson pairs
and $m$ neutral gauge bosons $\sNB =\PB,\PW^3$,
\beq
\Pe^+_\kappa \Pe^-_\kappa \rightarrow   \PW^+_{1,\rT}
\ldots \PW^+_{n,\rT}\PW^-_{1,\rT} \ldots \PW^-_{n,\rT} \sNB_{1,\rT}\ldots  \sNB_{m,\rT},
\eeq
the Born matrix element receives a factor $g_W=g_2$ for each
$\SUtwo$ gauge boson and a factor $g_B=g_1$ for each $\Uone$ gauge boson, 
and neglecting masses in the propagators we arrive at
\beq \label{symmbornampl} 
\tilde{\M}_0^{\Pep\Pem \PW^+_{1}\ldots \PW^-_{n}\sNB_1 \ldots \sNB_m }= \left(g_2^{2n}\prod_{k=1}^m g_{\sNB_k}\right)  \tilde{A}_0^{\Pep\Pem  \PW^+_{1}\ldots \PW^-_{n}\sNB_1 \ldots \sNB_m}.
\eeq 
The coupling-constant renormalization gives
\beq
\de^\pre=\left.\left(2n\frac{\de g_2}{g_2}+ \sum_{k=1}^m\frac{\de g_{\sNB_k}}{g_{\sNB_k}}\right) \right|_{\mu^2=s} .
\eeq
In the broken phase, the charged gauge bosons remain pure $\SUtwo$
eigenstates, and only the neutral gauge bosons mix.
In the high-energy limit, if we neglect the gauge-boson masses in
propagators, we can decompose the Born matrix elements into the
symmetric amplitudes \refeq{symmbornampl} using the Weinberg rotation.
For the production of $m$ neutral gauge bosons ($n=0$), we find
\beq
\M_0^{\Pep\Pem \NB_1 \ldots \NB_m}= \tilde{\M}_0^{\Pep\Pem \sNB_1 \ldots \sNB_m}\left[\prod_{k=1}^m U^{-1}_{\sNB_k \NB_k}(\thw)\right]=
\tilde{A}_0^{\Pep\Pem \sNB_1 \ldots  \sNB_m}
\left[\prod_{k=1}^m U^{-1}_{\sNB_k \NB_k}(\thw) g_{\sNB_k} \right],
\eeq
where sums over $\sNB_1 \ldots \sNB_m$ are implicitly understood, and the renormalization of coupling constants and mixing angle gives
\beqar
\lefteqn{\de^\pre \M^{\Pep\Pem \NB_1 \ldots \NB_m}=}\quad\nl
&=& \tilde{A}_0^{\Pep\Pem \sNB_1 \ldots \sNB_m} \sum_{l=1}^m
\left\{\left(\de U^{-1}_{\sNB_l \NB_l}(\thw) g_{\sNB_l}+ U^{-1}_{\sNB_l \NB_l}(\thw) \de g_{\sNB_l}\right) \left[\prod_{k=1, k\neq l}^m U^{-1}_{\sNB_k \NB_k}(\thw) g_{\sNB_k} \right]\right\}\nl
&=& \sum_{l=1}^m \M_0^{\Pep\Pem \NB_1 \ldots \NB'_l \ldots \NB_m}
\left(U_{\NB'_l\sNB_l}(\thw) \de U^{-1}_{\sNB_l\NB_l}(\thw)  + U_{\NB'_l\sNB_l}(\thw) \frac{\de g_{\sNB_l}}{g_{\sNB_l}}U^{-1}_{\sNB_l\NB_l}(\thw)\right).
\eeqar
Therefore, the complete PR correction can be written as
\beqar \label{deRGgb}
\de^{\pre} \M^{\Pep\Pem  \PW^+_{1} \ldots \PW^-_{n}\NB_1 \ldots \NB_m} &=& 2n\de^\pre_{WW} \M^{\Pep\Pem \PW^+_{1} \ldots \PW^-_{n} \NB_1 \ldots \NB_m}\nl
&&{}+\sum_{k=1}^m \delta^\pre_{\NB'_k \NB_k} \M_0^{\Pep\Pem \PW^+_{1} \ldots \PW^-_{n} \NB_1 \ldots \NB'_k \ldots \NB_{m}}
\eeqar
with the correction factors
\beqar\label{dergVT}
\de^{\pre}_{WW}&=&\left.\frac{\de g_2}{g_2}\, \right|_{\mu^2=s}= -\frac{1}{2}\bew_{W}\ls +\de Z^\elm_e,\nl
\de^\pre_{\NB'\NB}&=&\left.\left\{\left[U(\thw) \de U^{-1}(\thw)\right]_{\NB'\NB}+\left(\frac{\de g}{g}\right)_{\NB'\NB}\right\}\, \right|_{\mu^2=s} 
\nl&=&
-\frac{1}{2}\left[\antikro_{\NB'\NB}\bew_{AZ}+\bew_{\NB'\NB} \right]\ls +\de Z^\elm_e\, \de_{\NB'\NB},
\eeqar
where we have used \refeq{WRMAT},  $U\de
U^{-1}=-\de U U^{-1}$, and the coupling-renormalization matrix
\beq \label{couplrenmat}
\left(\frac{\de g}{g}\right)_{\NB'\NB}:=U_{\NB'\sNB}(\thw) \,\frac{\de g_{\sNB}}{g_{\sNB}} \,U^{-1}_{\sNB \NB}(\thw)=-\frac{1}{2} \bew_{\NB'\NB}\lu+ \de Z^{\elm}_e \de_{\NB'\NB},
\eeq
generated by Weinberg rotation of the two gauge-coupling counterterms
\refeq{gCTs}. 

Comparing the PR and collinear 
SL contributions for transverse gauge bosons,
\refeq{dergVT}, \refeq{deccWT} and \refeq{deccVVT}, we
find that all 
symmetric-electroweak
logarithms are related to the $\beta$-function 
and cancel in the sum so that only large logarithms of
pure electromagnetic origin contribute to the complete SL corrections,
\beqar
\de^\cc_{WW}(\GB_\rT)+\de^\pre_{WW}&=&\de Z^\elm_e+ Q_\PW^2 \lemW ,\nl
\de^\cc_{\NB'\NB}(\GB_\rT)+\de^\pre_{\NB'\NB}&=& \de Z^\elm_e\, \de_{\NB'Z}\de_{\NB Z}.
\eeqar
Note that this cancellation between PR and collinear logarithms occurs
already in the symmetric basis and is a consequence of Ward
identities, like the identity between the electric charge and the
photonic FRC in QED. In the physical basis, both coupling and field
renormalization constants receive additional terms owing to mixing of
the neutral gauge bosons, but also these terms cancel. The same
relation holds for an arbitrary fermion-antifermion pair in the
initial state.

\section{Representation of $\SUtwo\times \Uone$ operators} 
\label{app:representations} 
Generators of the gauge group and various group-theoretical matrices used in the article are presented in detail. 
Our notation for the components of such matrices is 
\beq
M_{\varphi_i\varphi_{i'}}(\varphi),
\eeq
where the argument $\varphi$ represents a multiplet and fixes the
representation for the matrix $M$, whereas $\varphi_i$ are the
components of the multiplet. In this appendix we give explicit
representations for left- and right-handed fermions
($\varphi=f^L,f^R,\bar{f}^L,\bar{f}^R$), for gauge bosons
($\varphi=\GB$) and for the scalar doublet ($\varphi=\Phi$).  Where the
representation is implicit, the argument $\varphi$ is omitted. For the
eigenvalues of diagonal matrices we write
\beq
M_{\varphi_i\varphi_{i'}}=\de_{\varphi_i\varphi_{i'}}M_{\varphi_i}.
\eeq  

\subsection*{Symmetric and physical gauge fields and gauge couplings}
For gauge bosons we take special care of the effect of Weinberg
rotation (mixing). The symmetric basis $\tilde{\GB}_a=B,W^1,W^2,W^3$, is
formed by the $\Uone$ and $\SUtwo$ gauge bosons, which transform as a
singlet and a triplet, respectively, and quantities in this basis are
denoted by a tilde. The physical basis is given by the charge and mass
eigenstates $\GB_a=A,Z,W^+,W^-$. The physical charged gauge bosons,
\begin{equation}
W^\pm=\frac{W^1\mp\ri W^2}{\sqrt{2}},
\end{equation} 
are pure $\SUtwo$ states, whereas in the neutral sector the $\SUtwo$
and $\Uone$ components mix, and the physical fields $\NB=A,Z$  are
related to the symmetric fields $\sNB=B,W^3$ by the Weinberg rotation, 
\beq \label{weinbergrotation}
\NB=U_{\NB\sNB}(\thw)\sNB,\qquad  U(\thw)=\left(\begin{array}{c@{\;}c}\cw & -\sw \\ \sw & \cw \end{array}\right)
\eeq
with $\cw=\cos{\thw}$ and $\sw=\sin{\thw}$. In the on shell
renormalization scheme the Weinberg angle is fixed by
\beq \label{mixingangle}
\cw=\frac{\MW}{\MZ}.
\eeq

The gauge couplings are given by the generators of global gauge transformations \refeq{generators}. 
In the symmetric basis, they read
\beq
\tilde{I}^B=-\frac{1}{\cw}\frac{Y}{2},\qquad \tilde{I}^{W^a}=\frac{1}{\sw}T^a,\qquad a=1,2,3,
\eeq
where $Y$ is the weak hypercharge and $T^a$ are the components of the weak isospin.  In the physical basis we have
\beq
I^A=-Q,\qquad I^Z=\frac{T^3-\sw^2 Q}{\sw\cw},\qquad
I^\pm=\frac{1}{\sw}T^\pm=\frac{1}{\sw}\frac{T^1\pm\ri T^2}{\sqrt{2}}
\eeq
with $Q=T^3+Y/2$.

\subsection*{Casimir operators}
The  $\SUtwo$ Casimir operator is defined by
\beq
C=\sum_{a=1}^3(T^a)^2.
\eeq
Loops involving charged gauge bosons are often associated with the
product of the non-abelian charges
\begin{equation}
(I^W)^2:=\sum_{\sigma=\pm}\left[ I^\sigma I^{-\sigma} \right]=\left[\frac{C-(T^3)^2}{\sw^2}\right],
\end{equation}
and if one includes the contributions of neutral gauge bosons, one obtains the effective electroweak Casimir operator
\begin{equation}\label{CasimirEW} 
\cew:=\sum_{\GB_a=A,Z,W^\pm} I^{\GB_a}I^{\bar{\GB}_a}=\frac{1}{\cw^2}\left(\frac{Y}{2}\right)^2+\frac{1}{\sw^2}C.
\end{equation}
For irreducible representations (fermions and scalars) with isospin
$T_\varphi$, the $\SUtwo$ Casimir operator is proportional to the
identity and reads
\beq
C_{\varphi_i\varphi_{i'}}(\varphi)=\de_{\varphi_i \varphi_{i'}}C_\varphi,\qquad C_\varphi=T_\varphi[T_\varphi+1].
\eeq
For gauge bosons we have a reducible representation. In the symmetric basis $\tilde{C}(\GB)$ is a diagonal $4\times 4$ matrix
\beq \label{44diagonal}
\tilde{C}_{\tilde{\GB}_a\tilde{\GB}_{b}}=\de_{ab}\tilde{C}_{\tilde{\GB}_a},
\eeq
with  $\Uone$ and $\SUtwo$ eigenvalues
\beq \label{adjointCasimir1}
\tilde{C}_{B}=0,\qquad \tilde{C}_{W^a}=2.
\eeq
The transformation of a matrix like \refeq{44diagonal} to the physical
basis, yields a $4\times 4$ matrix with diagonal $2\times 2$ block
structure, \ie without mixing between the charged sector ($W^\pm$) and
the neutral sector ($\NB=A,Z$). In the neutral sector $C(\GB)$ becomes
non-diagonal owing to mixing of the $\Uone$ and $\SUtwo$ eigenvalues,
\beq  \label{adjointCasimir2} 
C_{\NB\NB'}= \left[U(\thw) \tilde{C} U^{-1}(\thw)\right]_{\NB\NB'}=
2\left(\begin{array}{c@{\;}c}\sw^2 & -\sw\cw \\ -\sw\cw & \cw^2 \end{array}\right),
\eeq
whereas in the charged sector it remains diagonal,
\beq  \label{adjointCasimir3} 
C_{W^\si W^{\si'}}=2\de_{\si\si'}.
\eeq

\subsection*{Explicit values for $Y$, $Q$,  $T^3$, $C$, $(I^A)^2$,
  $(I^Z)^2$, $(I^W)^2$, $C^\ew$, and $I^\pm$} 
Here we list the eigenvalues (or components) of the operators $Y$,
$Q$, $T^3$, $C$, $(I^A)^2$, $(I^Z)^2$, $(I^W)^2$, $C^\ew$, and
$I^\pm$, that have to be inserted in our general results. For incoming
particles or outgoing antiparticles the values for the particles have
to be used, for incoming antiparticles or outgoing particles the
values of the antiparticles.

\subsubsection*{Fermions}
The fermionic doublets $f^\kappa=(f^\kappa_+, f^\kappa_-)^\rT$
transform according to the fundamental or trivial representations,
depending on the chirality $\kappa=\rL,\rR$. Except for $I^\pm$, the
above operators are diagonal. For lepton and quark doublets,
$L^\kappa= (\nu^\kappa, l^\kappa)^\rT$ and $Q^\kappa=(u^\kappa,
d^\kappa)^\rT$, their eigenvalues are
\beq \label{Llept}
\renewcommand{\arraystretch}{1.5}
\begin{array}{c@{\quad}|@{\quad}c@{\quad}c@{\quad}c@{\quad}c@{\quad}c@{\quad}c@{\quad}c@{\quad}c@{\quad}} 
& {Y}/{2}& Q & T^3 &C &(I^A)^2&(I^Z)^2&(I^W)^2 & C^\ew\\ 
\hline
 \nu^{\rL},\bar{\nu}^{\rL} &\mp \frac{1}{2}  &0 & \pm \frac{1}{2} & \frac{3}{4}&  0  & \frac{1}{4\sw^2\cw^2} & \frac{1}{2\sw^2}  & \frac{1+2\cw^2}{4\sw^2\cw^2}  \\

 l^{\rL},\bar{l}^{\rL} &\mp \frac{1}{2} &\mp 1 & \mp \frac{1}{2} & \frac{3}{4}   & 1  &  \frac{(\cw^2-\sw^2)^2}{4\sw^2\cw^2} & \frac{1}{2\sw^2} & \frac{1+2\cw^2}{4\sw^2\cw^2}  \\

 l^{\rR},\bar{l}^{\rR}  & \mp 1  & \mp 1 & 0 & 0  & 1  & \frac{\sw^2}{\cw^2} & 0  & \frac{1}{\cw^2} \\

 u^{\rL},\bar{u}^{\rL} &\pm \frac{1}{6} & \pm \frac{2}{3} & \pm \frac{1}{2} & \frac{3}{4}  &  \frac{4}{9}  &  \frac{(3\cw^2-\sw^2)^2}{36\sw^2\cw^2} & \frac{1}{2\sw^2}   & \frac{\sw^2+27\cw^2}{36\cw^2\sw^2} \\

 d^{\rL},\bar{d}^{\rL} &\pm \frac{1}{6} & \mp \frac{1}{3} & \mp \frac{1}{2} & \frac{3}{4}  &  \frac{1}{9}  &  \frac{(3\cw^2+\sw^2)^2}{36\sw^2\cw^2} & \frac{1}{2\sw^2}  & \frac{\sw^2+27\cw^2}{36\cw^2\sw^2} \\

 u^{\rR},\bar{u}^{\rR}  & \pm \frac{2}{3}  & \pm \frac{2}{3} & 0 & 0  & \frac{4}{9}  &  \frac{4}{9}\frac{\sw^2}{\cw^2} & 0 & \frac{4}{9\cw^2}  \\

 d^{\rR},\bar{d}^{\rR}  & \mp \frac{1}{3}  & \mp \frac{1}{3} & 0 & 0  &  \frac{1}{9}  & \frac{1}{9}\frac{\sw^2}{\cw^2} & 0 & \frac{1}{9\cw^2}  \\

\end{array}
\eeq
For left-handed fermions, $I^\pm(f^\rL)$ have  the non-vanishing components
\beq \label{ferpmcoup}
I^{\si}_{f_{\si'}f_{-\si'}}(f^\rL)=-I^{\si}_{\bar{f}_{-\si'}\bar{f}_{\si'}}(\bar{f}^\rL)=\frac{\de_{\si\si'}}{\sqrt{2}\sw},
\eeq
whereas for right-handed fermions $I^\pm(f^\rR)=0$.

\subsubsection*{Scalar fields}
The symmetric scalar doublet,
$\Phi= (\phi^+,\phi_0)^\rT$, $\Phi^*= (\phi^-,\phi_0^*)^\rT$,
transforms according to the fundamental representation, and its
quantum numbers correspond to those of left-handed leptons
\refeq{Llept} with 
\beq
\phi^+ \leftrightarrow \bar{l}^\rL, \qquad \phi_0 \leftrightarrow \bar{\nu}^\rL, \qquad
\phi^- \leftrightarrow {l}^\rL, \qquad \phi_0^* \leftrightarrow {\nu}^\rL.
\eeq
After symmetry breaking
the neutral scalar fields are parametrized by the mass eigenstates
\beq \label{Higgschi}
\phi_0=\frac{1}{\sqrt{2}} (v+ H + \ri\chi).
\eeq
With respect to this basis $S=(H,\chi)$ the operators $Q,C,(I^\NB)^2$,
and $\cew$ remain unchanged, while $T^3$ and $Y$ become non-diagonal
in the neutral components
\begin{equation}
T^3_{SS'}=-\left(\frac{Y}{2}\right)_{SS'}=
-\frac{1}{2}\left(\begin{array}{c@{\;}c}0 & \ri \\  -\ri & 0 \end{array}\right)
,
\end{equation}
and
\beq \label{ZHcoup}
I^Z_{H\chi}=-I^Z_{\chi H}=\frac{-\ri}{2\sw\cw}.
\eeq
The $\PW^\pm$ couplings read
\beq \label{scapmcoup}
I^{\si}_{S\phi^{-\si'}}=-I^{\si}_{\phi^{\si'}S}=\de_{\si\si'}I^{\si}_S
\eeq
with
\beq \label{scapmcoupB}
I^\si_{H}:=-\frac{\si}{2\sw},\qquad
I^\si_{\chi}:=-\frac{\ri}{2\sw}.
\eeq
\subsubsection*{Gauge fields}
For transversely polarized external gauge bosons we have to use the
adjoint representation. In the symmetric basis the diagonal operators have eigenvalues 
\begin{equation}\label{gaugeeigenvalues}
\renewcommand{\arraystretch}{1.5}
\begin{array}{c@{\quad}|@{\quad}c@{\quad}c@{\quad}c@{\quad}c@{\quad}c@{\quad}c@{\quad}c@{\quad}c@{\quad}} 
& {Y}/{2}&Q & T^3 &C &(I^A)^2&(I^Z)^2&(I^W)^2 & \cew  \\
\hline 

 W^\pm  & 0  & \pm 1  & \pm 1 & 2   & 1  & \frac{\cw^2}{\sw^2} &\frac{1}{\sw^2}  & \frac{2}{\sw^2}   \\

 W^3  &  0  &  0  & 0  & 2  & 0  & 0 & \frac{2}{\sw^2}  & \frac{2}{\sw^2}  \\

 B  &  0  & 0  & 0 & 0  & 0  & 0 & 0  & 0  \\
\end{array}
\end{equation}
In the neutral sector, owing to the Weinberg rotation, the non-trivial
operators $\cew,C$ and $(I^W)^2$ become non-diagonal in the physical
basis $\NB=A,Z$, 
with components
\beq
\cew_{\NB\NB'}=\frac{1}{\sw^2}C_{\NB\NB'}=(I^W)_{\NB\NB'}^2=
\frac{2}{\sw^2}\left(\begin{array}{c@{\;}c} \sw^2 & -\sw\cw \\ -\sw\cw & \cw^2 \end{array}\right),
\eeq
whereas the trivial operators ${Y}/{2}=Q=T^3=(I^A)^2=(I^Z)^2=0$ remain unchanged.
In the physical basis the non-vanishing components of the $I^\pm$ couplings are
\beq \label{gaupmcoup}
I^{\si}_{\NB W^{-\si'}}=-I^{\si}_{W^{\si'}\NB}=\de_{\si\si'}I^\si_\NB
\eeq
with
\beq
I^\si_A=-\si,\qquad
I^\si_Z=\si\frac{\cw}{\sw}.
\eeq

\subsection*{Dynkin operator}
The group-theoretical object appearing in gauge-boson self-energies is
the Dynkin operator
\begin{equation}\label{Dew} 
\dew_{ab}(\varphi):=\Tr\left\{{I}^{\bar{V}_a}(\varphi)I^{V_b}(\varphi)\right\}.
\end{equation}
The indices $a,b$ are those of the gauge group, and the trace is over
the isospin doublet for $\varphi =\Phi, f^\rL,
f^\rR$ and over the gauge group for $\varphi = \GB$. In the latter case
the Dynkin operator corresponds to the electroweak Casimir operator,
\beq 
\dew_{ab}(\GB)=\cew_{ab}(\GB).
\eeq
In the symmetric basis $\dsew$ is diagonal,
\beq 
\dsew_{ab}(\varphi)=\delta_{ab} \dsew_{a}(\varphi).
\eeq
The $\SUtwo$ and $\Uone$ eigenvalues of the fundamental
representation, $\varphi =\Phi ,f^\rL$, read 
\beq
\dsew_{B}(\varphi)=\frac{Y_{\varphi}^2}{2\cw^2},\qquad \dsew_{W}(\varphi)=\frac{1}{2\sw^2},
\eeq
while for right-handed fermions 
\beq
\qquad \dsew_{B}(f^\rR)=\frac{Y_{f^\rR_+}^2+Y_{f^\rR_-}^2}{4\cw^2},\qquad
\dsew_{W}(f^\rR)=0.
\eeq
In the physical basis  we have 
\beq
\dew_{W^\si W^{\si'}}(\varphi)=\de_{\si\si'} \dsew_{W}(\varphi)
\eeq
for the charged components, whereas in the neutral sector the $\Uone$
and $\SUtwo$ eigenvalues mix resulting in
\beq
\dew_{\NB\NB'}(\varphi)= \left[U(\thw) \dsew (\varphi)
  U^{-1}(\thw) \right]_{\NB\NB'}=
\frac{1}{2}\left(\begin{array}{c@{\;}c}1+Y_{\varphi}^2 &
    \frac{Y_{\varphi}^2\sw^2-\cw^2}{\sw\cw} \\[1ex]  
\frac{Y_{\varphi}^2\sw^2-\cw^2}{\sw\cw}  &  \frac{Y_{\varphi}^2\sw^4+\cw^4}{\sw^2\cw^2} \end{array}\right)
\eeq
for $\varphi =\Phi ,f^\rL$, and 
\beq
\dew_{\NB\NB'}(f^\rR) =\frac{Y_{f^R_+}^2+Y_{f^R_-}^2}{4\cw^2}\left(\begin{array}{c@{\;}c}\cw^2 & \cw\sw \\ \cw\sw  &  \sw^2 \end{array}\right).
\eeq
The explicit values of the  components of the Dynkin operator for the leptonic doublets (and for the scalar doublet) are
\begin{equation} \label{llDynkin} 
\renewcommand{\arraystretch}{1.5}
\begin{array}{c@{\quad}|@{\quad}c@{\quad}c@{\quad}c@{\quad}c@{\quad}} 
&   \dew_{AA} & \dew_{AZ} & \dew_{ZZ} & \dew_{W}\\ 
\hline
 L^{\rL}, \Phi& 1 & \frac{\sw^2-\cw^2}{2\sw\cw}  & \frac{\sw^4+\cw^4}{2\sw^2\cw^2} & \frac{1}{2\sw^2} \\
 L^{\rR}& 1 & \frac{\sw}{\cw}  & \frac{\sw^2}{\cw^2} & 0 \\
 L^{\rL}+L^{\rR}& 2 & \frac{3\sw^2-\cw^2}{2\sw\cw}  & \frac{3\sw^4+\cw^4}{2\sw^2\cw^2} & \frac{1}{2\sw^2} \\
\end{array}
\end{equation}
and for the quark doublets
\begin{equation}
\renewcommand{\arraystretch}{1.5}
\begin{array}{c@{\quad}|@{\quad}c@{\quad}c@{\quad}c@{\quad}c@{\quad}} 
&   \dew_{AA} & \dew_{AZ} & \dew_{ZZ} & \dew_{W}\\ 
\hline
 Q^{\rL}& \frac{5}{9} & \frac{\sw^2-9\cw^2}{18\sw\cw}  & \frac{\sw^4+9\cw^4}{18\sw^2\cw^2} & \frac{1}{2\sw^2} \\
 Q^{\rR}& \frac{5}{9} &\frac{5}{9} \frac{\sw}{\cw}  &\frac{5}{9} \frac{\sw^2}{\cw^2} & 0 \\
 Q^{\rL}+Q^{\rR}& \frac{10}{9} & \frac{11\sw^2-9\cw^2}{18\sw\cw}  & \frac{11\sw^4+9\cw^4}{18\sw^2\cw^2} & \frac{1}{2\sw^2} \\
\end{array}
\end{equation}

\subsection*{$\beta$-function coefficients}
In gauge-boson self-energies and mixing energies, the sums of
gauge-boson, scalar, and fermionic loops give the following
combination of Dynkin operators
\beq \label{betafunction}
\bew_{ab}:=\frac{11}{3}\cew_{ab}(\GB)-\frac{1}{3}\dew_{ab}(\phi)-\frac{2}{3}\sum_{f,i} \NCf \sum_{\la}\dew_{ab}(f^\lambda),
\eeq
which is proportional to the one-loop coefficients of the
$\beta$-function.
The fermionic sum runs over the generations $i=1,2,3$ for leptons and quarks $f=l,q$.  In the symmetric basis $\besw_{ab}$ is diagonal, and its eigenvalues
\beq
\besw_{B}= -\frac{41}{6\cw^2},\qquad 
\besw_{W}=\frac{19}{6\sw^2}
\eeq
describe the running of the $\Uone$ and $\SUtwo$ coupling constants. In the physical basis $\bew_{ab}$ remains diagonal in the charged sector
\beq
\bew_{W^\si W^{\si'}}=\de_{\si\si'}\bew_{W}
\eeq
with $\bew_{W}=\besw_{W}$, whereas the neutral components 
\beq  \label{vvbetafunction}
\bew_{\NB\NB'}= \left[U(\thw) \besw U^{-1}(\thw)\right]_{\NB\NB'} 
\eeq
are
\beqar \label{betarelations}
\bew_{AA}&=&\cw^2\besw_{B}+\sw^2\besw_{W}=-\frac{11}{3},\qquad
\bew_{AZ}=\cw\sw(\besw_{B}-\besw_{W})=-\frac{19+22\sw^2}{6\sw\cw},\nl
\bew_{ZZ}&=&\sw^2\besw_{B}+\cw^2\besw_{W}=\frac{19-38\sw^2-22\sw^4}{6\sw^2\cw^2}.
\eeqar
The $AA$ component determines the running of the electric
charge, and the $AZ$ component is associated with the running of the
weak mixing angle [\cf\refeq{chargerenorm} and \refeq{weinbergrenorm}].

 \newcommand{\vj}[4]{{\sl #1~}{\bf #2~}\ifnum#3<100 (19#3) \else (#3) \fi #4}
 \newcommand{\ej}[3]{{\bf #1~}\ifnum#2<100 (19#2) \else (#2) \fi #3}
 \newcommand{\vjs}[2]{{\sl #1~}{\bf #2}}

 \newcommand{\am}[3]{\vj{Ann.~Math.}{#1}{#2}{#3}}
 \newcommand{\ap}[3]{\vj{Ann.~Phys.}{#1}{#2}{#3}}
 \newcommand{\app}[3]{\vj{Acta~Phys.~Pol.}{#1}{#2}{#3}}
 \newcommand{\cmp}[3]{\vj{Commun. Math. Phys.}{#1}{#2}{#3}}
 \newcommand{\cnpp}[3]{\vj{Comments Nucl. Part. Phys.}{#1}{#2}{#3}}
 \newcommand{\cpc}[3]{\vj{Comp. Phys. Commun.}{#1}{#2}{#3}}
 \newcommand{\epj}[3]{\vj{Eur. Phys. J.}{#1}{#2}{#3}}
 \newcommand{\fp}[3]{\vj{Fortschr. Phys.}{#1}{#2}{#3}}
 \newcommand{\hpa}[3]{\vj{Helv. Phys.~Acta}{#1}{#2}{#3}}
 \newcommand{\ijmp}[3]{\vj{Int. J. Mod. Phys.}{#1}{#2}{#3}}
 \newcommand{\jetp}[3]{\vj{JETP}{#1}{#2}{#3}}
 \newcommand{\jetpl}[3]{\vj{JETP Lett.}{#1}{#2}{#3}}
 \newcommand{\jmp}[3]{\vj{J.~Math. Phys.}{#1}{#2}{#3}}
 \newcommand{\jp}[3]{\vj{J.~Phys.}{#1}{#2}{#3}}
 \newcommand{\lnc}[3]{\vj{Lett. Nuovo Cimento}{#1}{#2}{#3}}
 \newcommand{\mpl}[3]{\vj{Mod. Phys. Lett.}{#1}{#2}{#3}}
 \newcommand{\nc}[3]{\vj{Nuovo Cimento}{#1}{#2}{#3}}
 \newcommand{\nim}[3]{\vj{Nucl. Instr. Meth.}{#1}{#2}{#3}}
 \newcommand{\np}[3]{\vj{Nucl. Phys.}{#1}{#2}{#3}}
 \newcommand{\npbps}[3]{\vj{Nucl. Phys. B (Proc. Suppl.)}{#1}{#2}{#3}}
 \newcommand{\pl}[3]{\vj{Phys. Lett.}{#1}{#2}{#3}}
 \newcommand{\prp}[3]{\vj{Phys. Rep.}{#1}{#2}{#3}}
 \newcommand{\pr}[3]{\vj{Phys.~Rev.}{#1}{#2}{#3}}
 \newcommand{\prl}[3]{\vj{Phys. Rev. Lett.}{#1}{#2}{#3}}                       
 \newcommand{\ptp}[3]{\vj{Prog. Theor. Phys.}{#1}{#2}{#3}}                     
 \newcommand{\rpp}[3]{\vj{Rep. Prog. Phys.}{#1}{#2}{#3}}                       
 \newcommand{\rmp}[3]{\vj{Rev. Mod. Phys.}{#1}{#2}{#3}}                        
 \newcommand{\rnc}[3]{\vj{Revista del Nuovo Cim.}{#1}{#2}{#3}}                 
 \newcommand{\sjnp}[3]{\vj{Sov. J. Nucl. Phys.}{#1}{#2}{#3}}                   
 \newcommand{\sptp}[3]{\vj{Suppl. Prog. Theor. Phys.}{#1}{#2}{#3}}             
 \newcommand{\zp}[3]{\vj{Z. Phys.}{#1}{#2}{#3}}                                
 \renewcommand{\and}{and~}

\newpage

\end{document}